\documentclass[fleqn,usenatbib]{mnras}

\usepackage{mathptmx}

\usepackage[T1]{fontenc}

\DeclareRobustCommand{\VAN}[3]{#2}
\let\VANthebibliography\thebibliography
\def\thebibliography{\DeclareRobustCommand{\VAN}[3]{##3}\VANthebibliography}

\usepackage{graphicx}	
\usepackage{amsmath}	
\usepackage{amssymb}	

\title[An ERIS/NIX imaging survey]{An optimised survey strategy for the ERIS/NIX imager: searching for young giant exoplanets and very low mass brown dwarfs using the $K$-peak custom photometric filter}

\author[S.C. Dubber et al.]{
Sophie Dubber$^{1,2}$\thanks{E-mail: dubber@roe.ac.uk}, 
Beth Biller$^{1,2}$,
Mariangela Bonavita$^{1,2}$,
Katelyn Allers$^{3}$, 
Clémence Fontanive$^{4}$,
\newauthor
Matthew A. Kenworthy${^5}$,
Micka\"{e}l Bonnefoy${^6}$,
William Taylor$^{7}$
\\
$^{1}$SUPA, Institute for Astronomy, Royal Observatory, University of Edinburgh, Blackford Hill, Edinburgh EH93HJ, UK\\
$^{2}$Centre for Exoplanet Science, University of Edinburgh, Edinburgh, UK\\
$^{3}$Department of Physics and Astronomy, Bucknell University, Lewisburg, PA 17837, USA\\
$^{4}$Center for Space and Habitability, University of Bern, Gesellschaftsstrasse 6, 3012 Bern, Switzerland\\
$^{5}$Leiden Observatory, Leiden University, P.O. Box 9513, 2300 RA Leiden, The Netherlands \\
$^{6}$Univ. Grenoble Alpes, CNRS, IPAG, F-38000 Grenoble, France \\
$^{7}$STFC, UK Astronomy Technology Centre, Royal Observatory Edinburgh, Edinburgh, EH9 3HJ, UK \\
}

\date{Accepted XXX. Received YYY; in original form ZZZ}

\pubyear{2022}

\begin{document}
\label{firstpage}
\pagerange{\pageref{firstpage}--\pageref{lastpage}}
\maketitle

\begin{abstract}
We present optimal survey strategies for the upcoming NIX imager, part of the ERIS instrument to be installed on the Very Large Telescope (VLT). We will use a custom 2.2$\mu$m $K$-peak filter to optimise the efficiency of a future large-scale direct imaging survey, aiming to detect brown dwarfs and giant planets around nearby stars. We use the results of previous large scale imaging surveys (primarily SPHERE SHINE and Gemini GPIES) to inform our choice of targets, as well as improved planet population distributions. We present four possible approaches to optimise survey target lists for the highest yield of detections: i) targeting objects with anomalous proper motion trends, ii) a follow-up survey of dense fields from SPHERE SHINE and Gemini GPIES iii) surveying nearby star-forming regions and iv) targeting newly discovered members of nearby young moving groups. We also compare the predicted performance of NIX to other state-of-the-art direct imaging instruments. 

\end{abstract}




\section{Introduction} \label{sec:intro}

Tens of giant planet and brown dwarf companions have been discovered and extensively characterised by direct imaging in recent decades \citep[e.g][]{chauvin04,marois08,lagrange10}, using instruments and techniques designed to measure the light coming directly from companions. Direct imaging is complementary to other highly successful methods, for example generally targeting wider star-planet separations than techniques such as radial velocity or photometric transit. The newest generation of large direct imaging surveys, such as the SPHERE Infrared Survey for Exoplanets \citep[SHINE;][]{desidera21,langlois21,vigan21} and the Gemini Planet Imager Exoplanet Survey \citep[GPIES;][]{macintosh15,nielsen19}, are nearing the end of observations, and are beginning to shed light on the details of planet populations through their early statistical results \citep{nielsen19,vigan21}. Large direct imaging surveys of hundreds of targets can tell us about many aspects of exoplanet formation and evolution, including: the frequency of giant planets, brown dwarfs and binary systems \citep[e.g][]{montet14,lannier16,reggiani16,meyer18,baron19,fulton21,bonavita21}; the viability of disk and planet formation models \citep[e.g][]{janson11,janson12,rameau13,vigan17,nielsen19,vigan21}, and orbital dynamics of multi-planet systems \citep[e.g][]{konopacky16,wang18,nielsen20}. Successfully discovering new giant planets and brown dwarfs that are suitable targets for these kind of studies requires a solid understanding of where to look: which host stars to target and what contrasts and sensitivities are necessary. To know this in turn requires a firm understanding of the underlying populations of planets  \par

The first direct imaging surveys targeting large samples of stars were informed by planet populations derived from radial velocity results. Radial velocity surveys generally target a different part of parameter space to direct imaging, being most sensitive to high mass planets close to their host stars. Early studies of the first populations of objects discovered via radial velocity found that the planet occurrence rate was best fit by a rising power law in mass and orbital period \citep[e.g.][]{cumming08}. Due to the lack of giant planet detections at wider separations, these fits had to be extrapolated to inform direct imaging surveys. The number of planets was predicted to continue to increase with wider separations, implying that surveys targeting this parameter space would report many giant planet detections. Many of this generation of direct imaging surveys \citep[e.g][among others]{desidera15,chauvin15,biller13} ultimately reported null or fewer than expected detections, raising questions about the true underlying planet population. These results began to confirm what was suspected prior to the surveys: there are far fewer planets at wide separations than the original RV power law extension predicts. \par

Despite this, the most recent large direct imaging surveys, SPHERE SHINE \citep{chauvin17} and Gemini GPIES \citep{macintosh15,nielsen19} have each reported a handful of new detections, with detected planets and brown dwarfs covering a range of masses and separations \citetext{e.g. 51 Eri b, \citealp{macintosh15}; HIP 65426b, \citealp{chauvin17}; PDS 70b, \citealp{keppler18}}. We can use these results, in combination with the myriad of detections made using other methods, to begin to inform the next generation of surveys. When the latest large-scale imaging surveys were first designed, the extension of the distribution of RV-detected planets to predict survey yields remained the only available assumption, with many planets expected at wide separations. New studies \citep[e.g][]{fernandes19,fulton21} have in fact found a turn-over in giant planet frequency between $\sim$2 and 4 AU, with giant planets appearing scarce at wide separations. In particular, \citet{fulton21} find giant planet occurrences consistent with initial occurrence results from GPIES \citep{nielsen19}, implying that these may be the most accurate results to date for informing new imaging surveys. These state-of-the-art investigations show that understanding and correctly using the predictions for where we might find planets should be a key consideration when designing a survey. It is also clear that combining results from a variety of exoplanet detection techniques is crucial when deriving population statistics, as this allows us to study the demographics of host stars and exoplanets with dramatically different properties (such as planet mass, separation and stellar spectral type). For example, studies such as \citet{meyer18} have shown that microlensing detections, which probe primarily very low-mass planets around low-mass primaries, offer a unique addition to the sample. Making use of all of available detection techniques vastly improves the range of planet properties included in a dataset. \par

In this paper, we consider optimal survey designs for the upcoming Enhanced Resolution Imager and Spectrograph \citep[ERIS;][]{davies18} at the Very Large Telescope (VLT). ERIS was designed as a dual replacement/upgrade of SINFONI \citep{eisenhauer03,bonnet04} and NACO \citep{rousset02,lenzen98}, which have been providing near-IR adaptive optics (AO) capability on the VLT for over a decade. Composed of the SPIFFIER spectrograph and the NIX imager, ERIS will be a key instrument for the next generation of large-scale direct imaging surveys. NIX is a near- and mid-IR AO enabled imager, with a grating vector apodised phase plate (gvAPP) coronagraph \citep{otten17,boehle18,kenworthy18}, which operates from 2-5µm. It is equipped with two detectors, offering 24"x24" and 53"x53" fields of view, with pixel scales of 13 mas pix$^{-1}$ and 27 mas pix$^{-1}$ (with the former being comparable in resolution to SPHERE (11"x12.5", 12.25 mas pix$^{-1}$) , but with a much larger field of view). NIX will be complementary to observations undertaken with SPHERE and GPI (which operate in the near-IR): $L-$ and $M-$band follow up can be combined with $J,H,K$ detections to better distinguish between equilibrium and non-equilibrium chemistry models. Additionally, older, cooler planets and brown dwarf companions have typically been detected in the longer wavelength $L$-band \citep[e.g][]{vigan15,stone18,launhardt20}, which is also very competitive for detecting young protoplanets still embedded in circumplanetary material \citep[e.g][]{reggiani14,keppler18,launhardt20,jorquera21} that is very bright at wavelengths longer than 3$\mu$m \citep[e.g][]{eisner15,szulagyi19}. \par

We designed the $K$-peak filter, a 6$\%$ width custom filter at 2.2µm, specifically for the detection of YPMOs via their `spectral shape'. The $K$-peak filter has been installed in the NIX imager, and can be used with the coronagraphic capabilities of the instrument for near-IR imaging. The design of our custom filter was informed by the spectral shape of very low mass objects in the $K$-band, which is significantly different to the spectral shape of earlier type objects in the same wavelength range. By using a specific combination of filters ($K$-peak, IB2.42 and $H$2-cont), we can trace the spectral shape of an object, and use the calculated colours to begin to  characterise an observed object. \par

Candidate companions detected in the course of large direct imaging surveys require multiple observations over several epochs for confirmation, as it must be shown that the candidate companion shares common proper motion and is actually bound to the target star, as opposed to being a background interloper.  This significantly increases the amount of telescope time necessary to fully complete a direct imaging survey and makes it difficult to confirm companions around stars with very low proper motions. Additionally, the most crowded fields located in the Galactic plane can often have hundreds of candidate companions identified around a single host star \citep[e.g.][]{vigan21}. Again, refuting or confirming each of these based on their proper motion could be very observationally intensive. An alternative option, obtaining spectroscopy for every candidate to check for an appropriate young planetary mass object (YPMO) spectrum as opposed to a background M star, is similarly unrealistic in scope. \par

The custom $K$-peak filter and spectral shape technique offer an option for drastically narrowing the number of candidate companions of interest in each field without proper motion follow-up, by providing crucial diagnostic information that can be used to approximately classify a candidate companion as a bonafide very low mass object or a background contaminant. Follow-up observations can then confirm this initial characterisation. Previous surveys have used a similar approach, with specific combinations of photometric filters used to aid characterisation: the SPHERE SHINE survey makes use of narrowband filters optimised to detect methane features \citep[e.g.][]{bonnefoy18}. Our technique builds on the same ideas, and in many cases could be used in combination with photometry from previous surveys to allow more robust characterisation than previously possible. \par 

We expect that our custom $K$-peak filter and spectral shape technique could be used to carry out a large survey for planetary-mass companions. As previous surveys have demonstrated, careful target selection is crucial to maximise resulting yields. We can learn from the model assumptions and target selection criteria of the SHINE and GPIES surveys, and also from the wealth of survey data that has been published since. This paper covers the design and target selection process for such a future survey. In Section \ref{sec:context}, we demonstrate the need for the spectral shape technique, by examining archival data. In Section \ref{sec:customfilter}, we discuss the design of the custom $K$-peak filter and show the diagnostic capabilities of the spectral shape technique. In Section \ref{sec:surveys}, we present multiple survey design options, and in Section \ref{sec:disc} we weigh the advantages and disadvantages of each observational approach, using the latest planet population models. 

\section{Previous Surveys and the Proper Motion Problem} \label{sec:context}

\subsection{Previous Surveys and Archival Imaging Coverage} \label{sec:context_archive}

\begin{figure*}
    \centering
    \includegraphics[width=\textwidth]{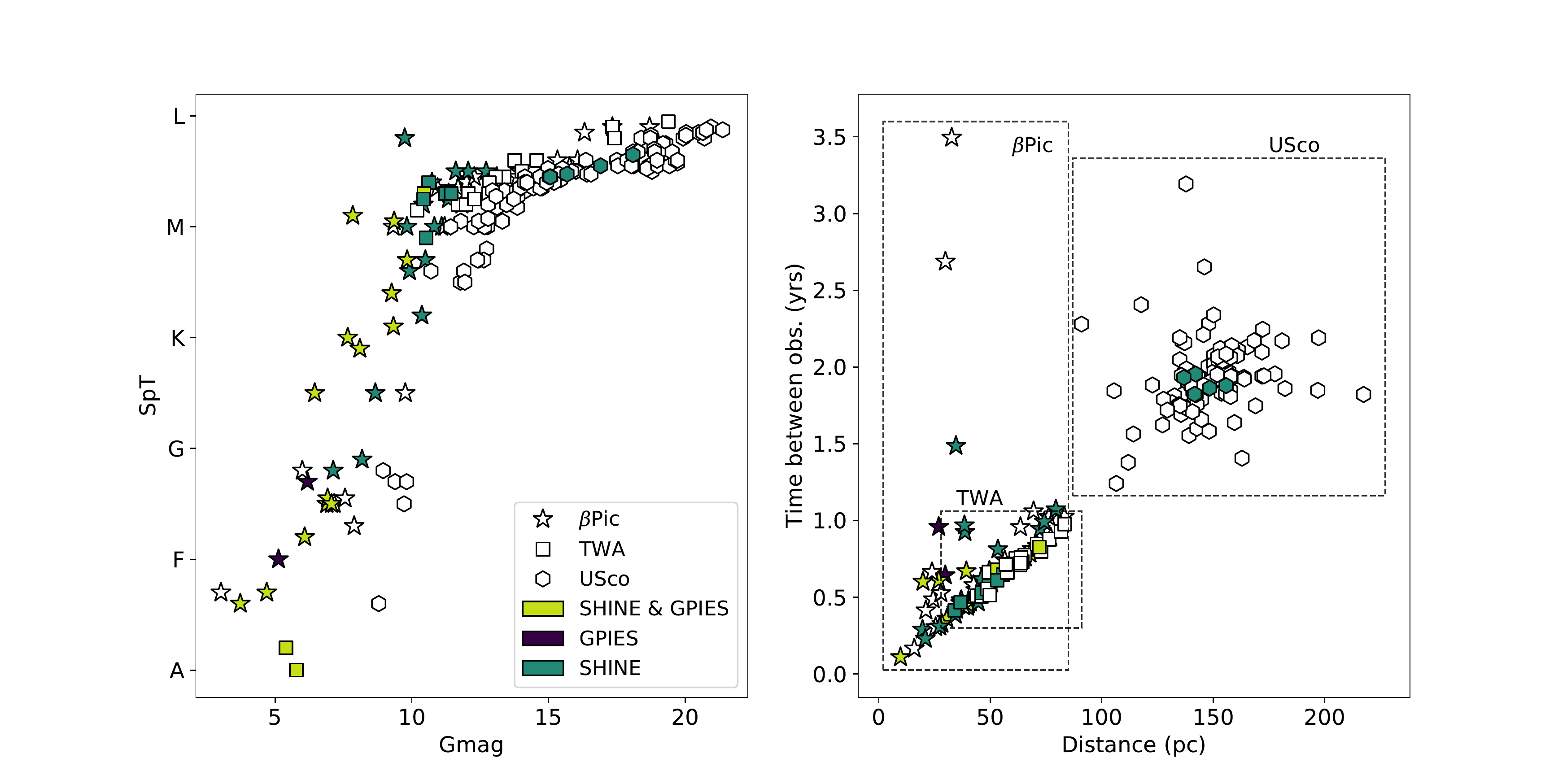}
    \caption{Left: $G$-magnitude vs Spectral Type for objects in the Upper Scorpius (hexagons), $\beta$ Pictoris (stars) and TW Hya (squares) membership lists, colour-coded based on archival imaging data. Objects with SPHERE/SHINE observations are highlighted in blue, objects with GPIES observations are highlighted in purple, and objects previously targeted by both surveys are highlighted in green. Right: Time required between observations to measure an on-sky movement of 50mas, as a function of distance to target star. Clustering of each region is identified by dashed boxes.}
    \label{fig:disc}
\end{figure*}

When planning a new survey of any kind, one must consider the previous archival coverage of various regions, and which have been most frequently targeted. There are two arguments to be made here. Firstly, archival imaging data can be extremely useful for increasing the observational baseline for a specific object. As a result, targeting regions that have been routinely observed in the past can actually prove very useful for confirming candidates. Secondly however, one should consider the likely remaining yield in a region if it has already been observed many times. It is highly unlikely that the most extensively surveyed have no planetary companions left to be discovered, but equally, the likelihood of discovering something new is somewhat reduced, as most remaining companions are likely below our current mass sensitivity.

Large-scale direct imaging surveys have been underway for the last two decades. As such, the scope of archival imaging data is broad. In this work, when considering possible observational approaches for a survey using NIX, we will focus primarily on star-forming regions and nearby young moving groups as example targets. Young, nearby moving groups are youthful associations of stars, usually no more than a couple of hundred parsecs from the sun. Young star-forming regions are similar in many respects, usually further away but generally youthful - the distinguishing factor being that they contain signs of very recent or ongoing star-formation (such as nebulosity and high-mass OB stars). Youth is a key selection criteria in direct imaging surveys that operate in the infrared (IR), as directly imaged planets cool and dim with age: it is easiest to detect them when they are young and at their brightest. There has been an historic favouring of some regions over others: the earliest imaging surveys all targeted the closest stars \citep[e.g][among others]{chauvin03,masciadri05,biller07}, due to the limited instrumental capabilities. These were followed by many thorough surveys of nearby young moving groups \citep[e.g][]{chauvin10,biller13,brandt14}, and there was generally far less focus on more distant star-forming regions. In this work, we will consider both moving groups and star-forming regions as potential targets for NIX.

In Section \ref{sec:surveys}, we will present results for possible surveys of the Upper Scorpius star forming region. Upper Scorpius is a representative example of star-forming regions that we could image with NIX. Part of the Scorpius-Centaurus Association (the closest OB association to the sun), it is located at 145 pc \citep{deBruijne97}, and is thought to have an age of 5-10 Myr \citep{pecaut12,pecaut16,david19}. It is the youngest of the three subgroups of Scorpius Centaurus (comprised of Upper Scorpius, Upper Centaurus–Lupus and Lower Centaurus–Crux) and is thought to contain $\sim$ 2500 members, 75\% of which have masses $\lesssim$ 0.6 M$_{\odot}$ \citep{preibisch08}. The low mass population in Upper Scorpius has been well explored in recent years, with the mass function below the stellar/substellar limit reasonably well-defined \citep{luhman20}. Many other star forming regions would be similarly suitable targets for a NIX survey, including Ophiuchus \citep[$\sim$130 pc;][]{ortiz18,canovas19}, Chamaeleon \citep[$\sim$180--200 pc;][]{voirin18,roccatagliata18} and Lupus \citep[$\sim$150--200 pc;][]{comeron08,gaia18}. \\

To obtain a list of likely Upper Scorpius members and their properties, we used the compilation presented in \citet{luhman18}, who reviewed the previous member lists in the literature and also obtained new spectroscopic observations to characterise hundreds of Upper Scorpius members. Starting from this list, we applied a magnitude cut based on the predicted performance of the NIX imager. According to a preliminary version of the ERIS manual, the limiting magnitude in $R$-band (the wavelength of operation for the wavefront sensing system) will be approximately 14 mags. We used a cautious lower limit of $R$=12 to allow for all degrees of observational conditions. We cross-matched the Upper Scorpius member list with the USNO-B all-sky catalogue \citep[][]{monet03}, as every member has a corresponding $R-$band magnitude from this survey. Due to the use of photographic plates in the USNO-B survey, the average photometric accuracy is 0.3 mag. We applied the $R$=12 to the Upper Scorpius member list, leaving us with a sample of 141 targets suitable for NIX observations.

Additionally, we will consider two moving groups as potential NIX targets in this work, TW Hya and $\beta$ Pictoris. A recent summary of young moving groups is given in \citet{gagne18a}, who detail compilations of members identified using GAIA-Tycho data \citep{hog00,gaia16a,gaia16b}. For specific member lists, we used the more recent work of \citet{carter21}, who present detailed compilations of the $\beta$ Pictoris and TW Hya moving groups, based on \citet{gagne18a}. As discussed in detail in \citet{carter21}, these moving groups are ideal targets for a general direct-imaging survey. $\beta$ Pictoris is located at $\approx$ 35 pc from the sun, and has an estimated age of 24$\pm$3 Myr \citep{bell2015}. The closest young group to the sun, it is an optimal target for an NIX survey: the young age corresponds to very bright young giant planets and brown dwarfs, reducing the contrast required for successful observations, and the close distance allows for detections of companions at closer separations to their target stars, a variable dictated entirely by the inner working angle of the instrument. Furthermore, moving groups in general have better defined age estimates than unassociated nearby stars, allowing for more precise mass estimates of detected companions. The second young moving group that we use as an example for NIX is the TW Hya association \citep{kastner97,gagne18a}. It is located at a slightly larger distance than $\beta$ Pictoris, at $\approx$ 60 pc, but is younger with an estimated age of 10$\pm$3 Myr \citep{bell2015}. Consequently, TW Hya presents the same observational advantages as $\beta$ Pictoris for a direct-imaging survey: being even younger, we expect lower mass planets to have sufficient luminosities for detection (due to their more recent formation), although at slightly larger separations due to the larger distance to TW Hya. While we focus on these two moving groups in this work, many others, including Octans, Columba and Carina \citep{gagne18a}, would be similarly suitable for a NIX survey. 

Using these three target lists, we investigated the scope of archival imaging data. We first queried the ESO and Gemini archives looking for SHINE and GPIES data, respectively. We then broadened the ESO archive search to look for matches with any previous direct imaging survey. 

Figure \ref{fig:disc} summarises the results of our archive search for SHINE and GPIES data, highlighting which host stars in our target lists have been observed previously. In Upper Scorpius, only 4.2\% of objects in our target list have SHINE observations, either published in the F150 sample papers \citep{vigan21,desidera21,langlois21} or observed in the remainder of survey time. None of the host stars in the list have been observed by GPIES. As expected, the coverage in the two moving groups is much higher. In particular, 53\% of objects in our $\beta$ Pictoris compilation have SHINE observations, and 27\% have been targeted by GPIES. 33\% of TW Hya objects have SHINE observations, and 10\% have been observed by GPIES. 

Figure \ref{fig:disc} shows the distribution of host stars in our samples in magnitude-spectral type space. We plot $G$-band magnitude, as the majority of stars have a detected magnitude in this filter, and colour-code based on available archival observations (SHINE only, GPIES only, both or neither).
The shape of the markers also indicates the membership of each object. In this distribution, we can see a clear bias in the target selection of these large surveys. While they have sampled a large range of spectral types, from early A to late M, the focus has clearly been on the brightest stars in each region. This is a consequence of instrument sensitivity and performance, and it is likely that NIX will be able to target some of the fainter objects plotted here (see Section \ref{sec:surveys}). As mentioned previously, the emphasis of surveys to date has been on young moving group stars: the majority of Upper Scorpius targets points are unfilled.  

Considering all archival image data that is presented in the ESO archive (including all observations categorised as `imaging', which includes historic data that may not be high contrast or coronagraphic), we see a similar overall picture. The Upper Scorpius targets have the lowest coverage in terms of archival data: but still at the 64\% level, meaning there could well be an existing long baseline of observations for any stars that are targeted in the future. The $\beta$ Pictoris and TW Hya moving group targets have 92\% and 100\% coverage in the ESO archive, respectively, again demonstrating the bias towards moving group members in previous surveys. 

Beyond the SHINE and GPIES surveys, state-of-the-art instruments are currently being used for numerous, slightly smaller direct imaging surveys \citep[e.g][]{launhardt20}, often targeting specific regions. Some of these have targeted Upper Scorpius, or the larger Scorpius-Centaurus (Sco-Cen) region in which it sits. BEAST \citep{janson21}, one such survey using SPHERE, is observing 85 B-type stars in Sco-Cen, 11 of which are located in Upper Scorpius. First results from this survey have proven successful, with 6 previously unknown stellar companions detected. Another ongoing Sco-Cen SPHERE survey, YSES \citep{bohn2020a}, aims to find planetary companions to 70 K-type stars in Lower Centaurus Crux (LCC). Two planetary systems have been discovered to date by the ongoing survey \citep{bohn2020a,bohn2020b,bohn21}. The current success of surveys focused on Sco-Cen subgroups is a promising sign for a future NIX survey of Upper Scorpius, and indicates that there are likely many planetary companions still to be found.  \par

The first large $L'$ surveys have also been undertaken in recent years ($\gtrsim 100$ targets), and can give us an insight into the potential results of a NIX survey in $L'$. \citet{stone18} present the LEECH survey, a large $L'$ survey that observed 98 nearby B--M type stars. They report one new low-mass companion, and are also able to place tight constraints of giant planet frequencies. The NACO-ISPY survey \citep{launhardt20} is observing 200 young stars, selected because they host debris or protoplanetary disks. Results from the first 2.5 years of ISPY have been released, with multiple new low-mass stellar companions reported \citep{cugno19,launhardt20}, and imaging of multiple disks in $L'$ for the first time. The process of target selection for this survey differs from how we would approach a NIX $L'$ survey, but it demonstrates the potential of $L'$ observations and the advantages of using longer wavelengths to image giant planets. Following the same trend seen when looking at most of the available archival data, there is no coverage of the targets in our Upper Scorpius list by these two $L'$ surveys. There is also minimal coverage of the TW Hya and $\beta$ Pictoris moving groups, likely because the target choice in these cases is motivated by factors other than proximity. These studies, along with earlier surveys targeting fewer host stars \citep[e.g.][54 nearby Sun-like stars]{heinze10}, demonstrate that longer-wavelength direct imaging surveys are just as valuable as the more common $J,H,K$ studies, and that instruments such as NIX, that offer $L'$ capabilities, ought to be extensively utilised.

\subsection{Proper motion considerations} \label{sec:context_pm}

\begin{figure*}
    \centering
    \includegraphics[width=\textwidth]{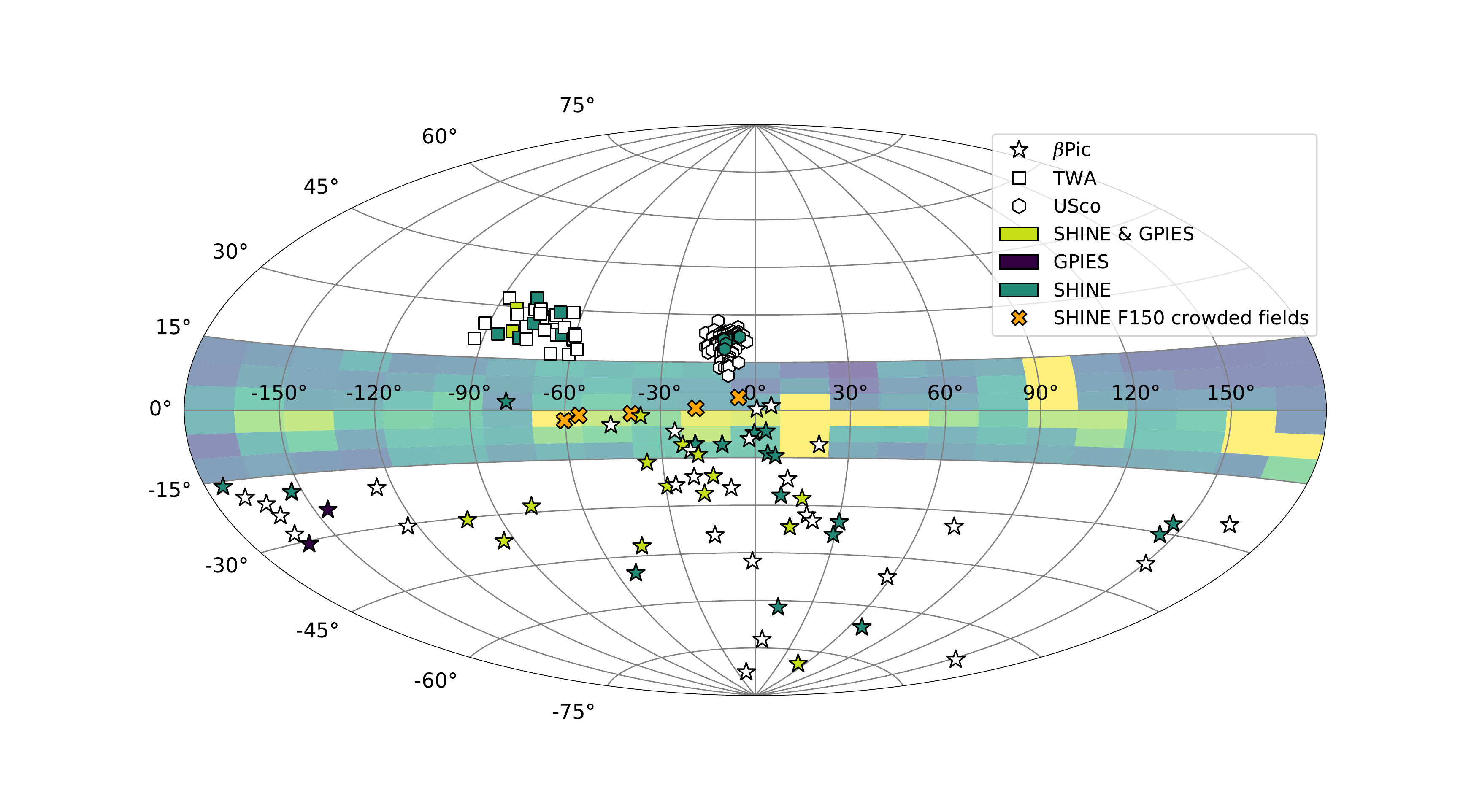}
    \caption{Sky positions of Upper Scorpius, $\beta$ Pictoris and TW Hya targets, projected onto a Galactic coordinate grid. Colour-coded based on archival imaging data: objects with SPHERE/SHINE observations are highlighted in blue, objects with GPIES observations are highlighted in purple, and objects previously targeted by both surveys are highlighted in green. The density of sources in the Galactic plane ($\pm15^{\circ}$) is also shown, with yellow corresponding to the highest density of sources and purple to the lowest.}
    \label{fig:gal}
\end{figure*}

As discussed in Section \ref{sec:intro}, a major goal of the $K$-peak filter and spectral shape technique is to provide the necessary information to facilitate rough characterisation of candidates from photometry alone, and enable the efficient removal of obvious contaminant objects. This is of particular importance for host star targets in field with low common proper motion.
The right panel of Figure \ref{fig:disc} demonstrates the population of targets for which our technique could prove very useful. Shown here is the distance to each host star target in the three membership lists considered in this paper, plotted against the time (in years) required between observations to confirm common proper motion. This was calculated based on the total proper motion of each star, and the assumption that a minimum of 50 mas of on-sky movement would be required to confidently confirm or refute a candidate companion. As expected, the different regions (located at distinct distance ranges) are obvious from the clustering in the diagram (and shown explicitly by the black dashed boxes). Again, the colour of the markers indicates objects with SHINE or GPIES archival data. \par

Figure \ref{fig:disc} directly demonstrates the difficulties with common proper motion follow up. Previous surveys have tended to focus on the closest young moving groups, which require a shorter time baseline to confirm any candidate companions - this can clearly be seen by the clustering of coloured points in the lower left of parameter space. In crowded, distant fields where many candidate companions are identified by current techniques and filter sets, follow-up surveys could be lengthy if a baseline of 3 years is required between observations. With the improved diagnostic capabilities of the $K$-peak filter and spectral shape technique (see Section \ref{sec:customfilter}), we aim to reject far more candidate companions using just first epoch photometry. Increased confidence that the remaining candidates are objects of interest for the survey, thus improving the likelihood of a high overall yield, will make extremely long baseline follow-up a more appealing option. \par

\subsection{Crowded Fields}

A final consideration when picking a region to target for a survey is its position in the sky with respect to the Galactic plane, bulge and other crowded areas. Figure \ref{fig:gal} shows the sky positions of the members of the three regions we consider in this work, with the same colour coding as used in the previous section. The clustering and small spatial extent of Upper Scorpius is clear here, due to its large distance. The contrast between the moving group targets is also striking - members of $\beta$ Pictoris are located across the sky, whereas TW Hya is more compact. A key detail to consider here is the location of the Galactic plane in Figure \ref{fig:gal} (Galactic latitudes $\pm15^{\circ}$). The low-resolution map covering these coordinates shows the source density across a grid of lines of sight in the Galactic plane and bulge (obtained using Simbad source counts), where yellow indicates the regions with the highest source density, and purple the lowest. The crowded bulge region is obvious around 0$^{\circ}$, and the source density elsewhere in the plane is comparatively low, but will still be far higher than areas of sky outside the Galactic plane.

We can see that many possible Upper Scorpius targets overlap with the upper Galactic plane, meaning any observations of host stars would require imaging very crowded fields. Upper Scorpius targets that have been imaged by the SHINE campaign (blue hexagons) lie outside the Galactic plane area, likely for this very reason. The lack of follow-up characterisation of so many companions is explained by their location, all being in the center of the Galactic plane. Furthermore, the Galactic bulge extends beyond the plane shown here, and is far brighter and more densely crowded than any other part of the sky - this further explains why the majority of Upper Scorpius targets are lacking follow-up. By increasing our ability to remove numerous contaminants with a single epoch of observations, the spectral shape technique could be used to target Upper Scorpius members that lie along the Galactic plane line-of-sight. \par

\section{$K$-peak custom filter} \label{sec:customfilter}

\begin{figure*}
    \centering
    \includegraphics[width=\textwidth]{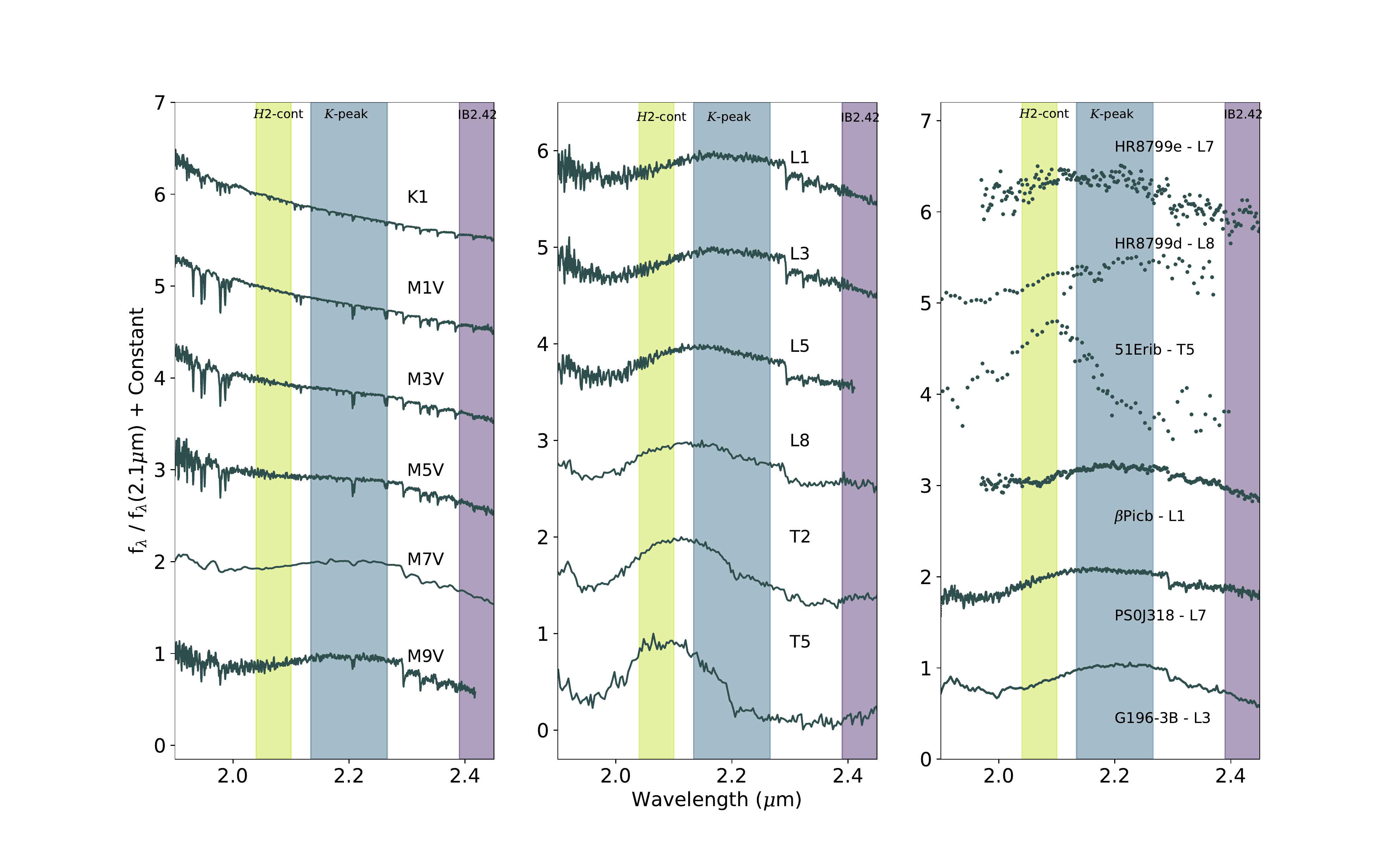}
    \caption{Left and Center: Spectral sequence for K-T spectral templates, showing the change in the spectral shape across the $K$-band. The three filters used for the spectral shape technique are highlighted: $H$2-cont (green), $K$-peak (blue) and IB2.42 (purple). Right: $K$-band spectral of well-known directly imaged exoplanets and brown dwarfs. References for spectral data: G196-3B, \citet{burgasser14} (SpeX prism library); PS0J 318, \citet{liu13}; $\beta$ Pictoris b, \citet{gravity20}; 51 Eri b, \citet{rajan17}; HR 8799 d, \citet{greenbaum18}; HR 8799 e, \citet{gravity19}.}
    \label{fig:spectra}
\end{figure*}

\subsection{Filter Motivation} \label{sec:customfilter_motivation}

In the previous sections, we have shown the difficulty in balancing follow-up time with large numbers of unconfirmed candidates companions from single-epoch photometry. This follow-up dilemma was the motivation for designing the $K$-peak filter: can we optimise the observing time required to determine the nature of a target? One answer is to use a carefully chosen combination of photometric filters. Using only photometry, we can calculate colours that contain information about the type of object being observed, allowing approximate characterisation with single-epoch photometry. Past works \citep[e.g][]{najita00,allers20} have shown that custom photometric filters can be used to greatly improve the confirmation rate of photometrically selected candidate low mass brown dwarfs. In previous work \citep{allers20,jose20,dubber21}, we used a custom filter centred on the deep 1.45µm feature present in YPMOs to distinguish between them and background sources. In the 2-5µm range covered by NIX, such water features are far less dominant, and there are strong telluric features across some of this range that would make a similar 'water' technique difficult to use. Instead, we use the differing spectral shape in $K$-band of very low mass brown dwarfs when compared to earlier spectral type stars. This can be seen in the sample of spectra shown in Figure \ref{fig:spectra}. By locating filters at key spectral points for defining the overall shape of the spectra, the extracted colour information can be used for direct characterisation. Also shown in Figure \ref{fig:spectra} are spectra of well-studied brown dwarfs and exoplanets, discovered via direct imaging: 51 Eri b \citep{macintosh15}, $\beta$ Pictoris b \citep{lagrange10}, PSOJ-318 \citep{liu13}, G196-3B \citep{rebolo98} and HR 8799d and e \citep{marois08,marois10}. References for the spectral data plotted are detailed in the caption of Figure \ref{fig:spectra}. These spectra demonstrate the general variety in the spectral shapes of objects that have been detected via direct imaging previously, but also the similar features in the highlighted filter windows. 

\subsection{Filter Design and Diagnostic Properties} \label{sec:customfilter_design}

The choice of waveband was the first consideration when designing the custom filter. The wavelength coverage of NIX (2-5µm) allows either $K$- or $L$- band as the two possible options for a custom filter.  Objects of interest (YPMOs) tend to be brighter in $L'$ than $K$, but the background level is also much higher, a crucial consideration for imaging companions. Additionally, the spectra of objects tend to be flatter and more featureless in $L'$ than $K$. Based on these factors, the $K$-band was chosen as the optimal band for our NIX custom filter. 

Next, we considered the question of width and positioning of the filter. One certainty dictated by the design of the instrument is that the coronagraph should not be used with wide filters. Doing so would lead to spectra rather than point source images due to the high spatial spread of the light. Consequently, we considered medium width filters centred at different wavelengths in $K$-band, in combination with the standard $H$2-cont, IB2.42 and IB2.48 filters installed in NIX. The $H$2-cont filter has a central wavelength of 2.07$\mu$m, and a width of $\Delta\lambda$/$\lambda \approx 3\%$. The IB2.42 filter has a central wavelength of 2.42$\mu$m, and a width of $\approx 2.5\%$.

\begin{figure}
    \centering
    \includegraphics[width=\columnwidth]{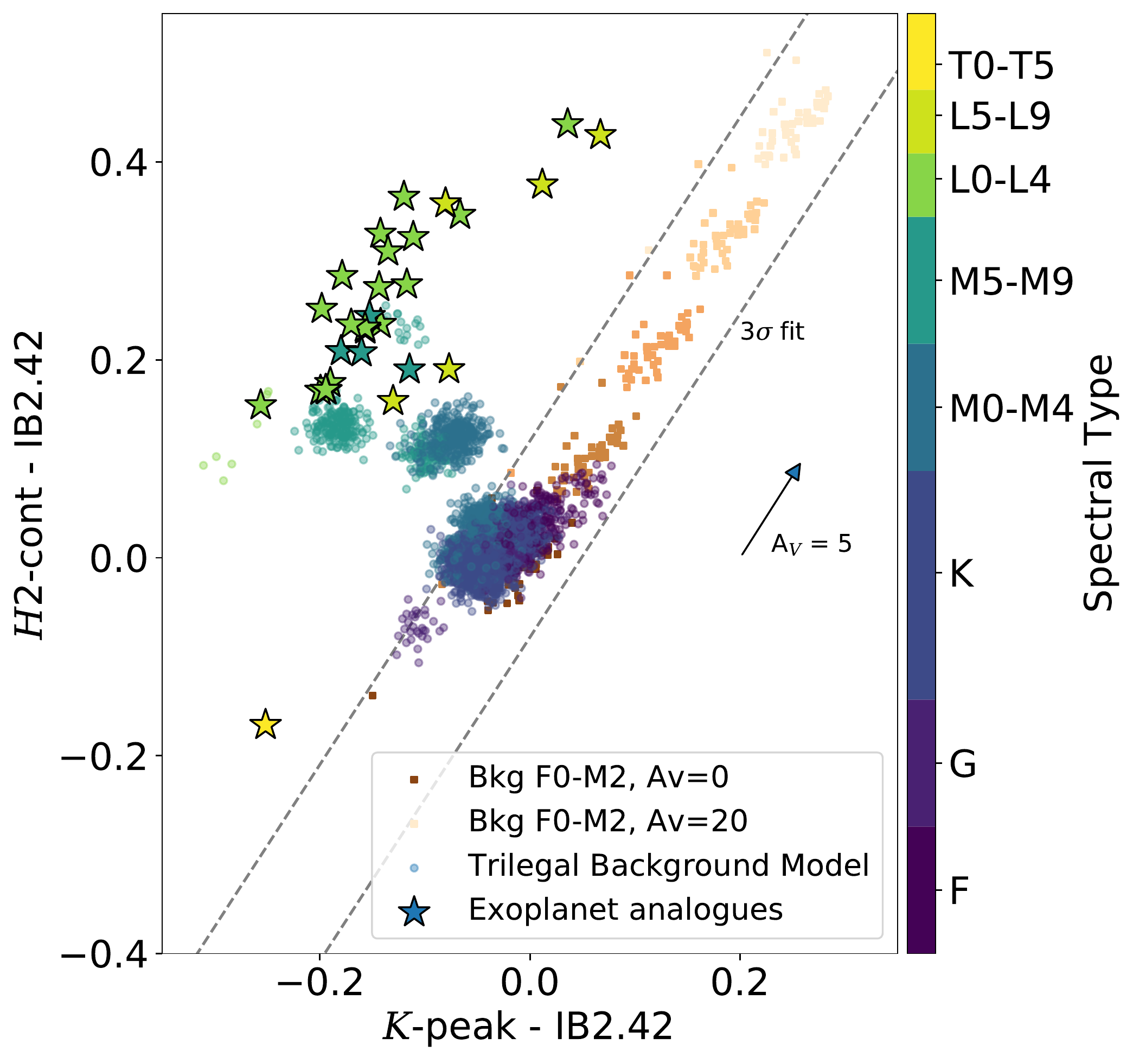}
    \caption{Colour-colour diagram for 3-filter spectral shape method. NIX filters $H$2-cont, $K$-peak and IB2.42 are used to characterise young, late-type exoplanet analogues (stars) from background population (circles and squares). Objects returned for Trilegal Galactic model along the $\beta$ Pictoris line-of-sight are plotted (circles), with the point colour indicating the spectral type of each object. Also shown is a reddened sequence of F0-M2 standard stars (coloured squares), with $A_{\text{V}}$ ranging from 0-20, in increments of $A_{\text{V}}$=5. Dashed lines show a 3$\sigma$ fit to this F0-M2 sequence. The arrow plotted shows the $A_{\text{V}}$ = 5 extinction vector.}
    \label{fig:nix_colcol}
\end{figure}

The properties of our final custom filter, as well as the standard IB2.42 and $H$2-cont filters, are shown in Figure \ref{fig:spectra}. The left and middle panels of Figure \ref{fig:spectra} also show a sequence of $K$-band spectra, starting with background stars (K1-M5) and moving through to ultracool brown dwarfs/ YPMOs (M7-T5). The variation in spectral shape with spectral type is clear in this wavelength range, with earlier spectral type objects having overall flatter spectra  than the late-M - mid-T objects, and a downward slope across the full $K$-band. Later-type objects have complex spectral gradients and features, echoed in the spectra of direct imaging detected planets and brown dwarfs (right panel). This was the basis for the spectral shape technique and custom filter design. Using two standard narrow filters also in the $K$-band, $H$2-cont and IB2.42 (highlighted in green and purple in Figure \ref{fig:spectra}), we considered different custom filter designs that would allow us to calculate colours that trace the evolving shape of the spectrum. To judge the effectiveness in distinguishing between populations of each possible filter design, we plotted colour-colour diagram for a range of possible colours and for a range of simulated object spectra, with [custom] - IB2.42 vs $H$2-cont - IB2.42 in each case. Through this iterative process, we chose the final custom filter characteristics that best distinguished between late spectral type objects (planets and brown dwarfs) and reddened earlier spectral type interlopers. Figure \ref{fig:nix_colcol} shows the colour-colour diagram for the final design.

\begin{figure}
    \centering
    \includegraphics[width=\columnwidth]{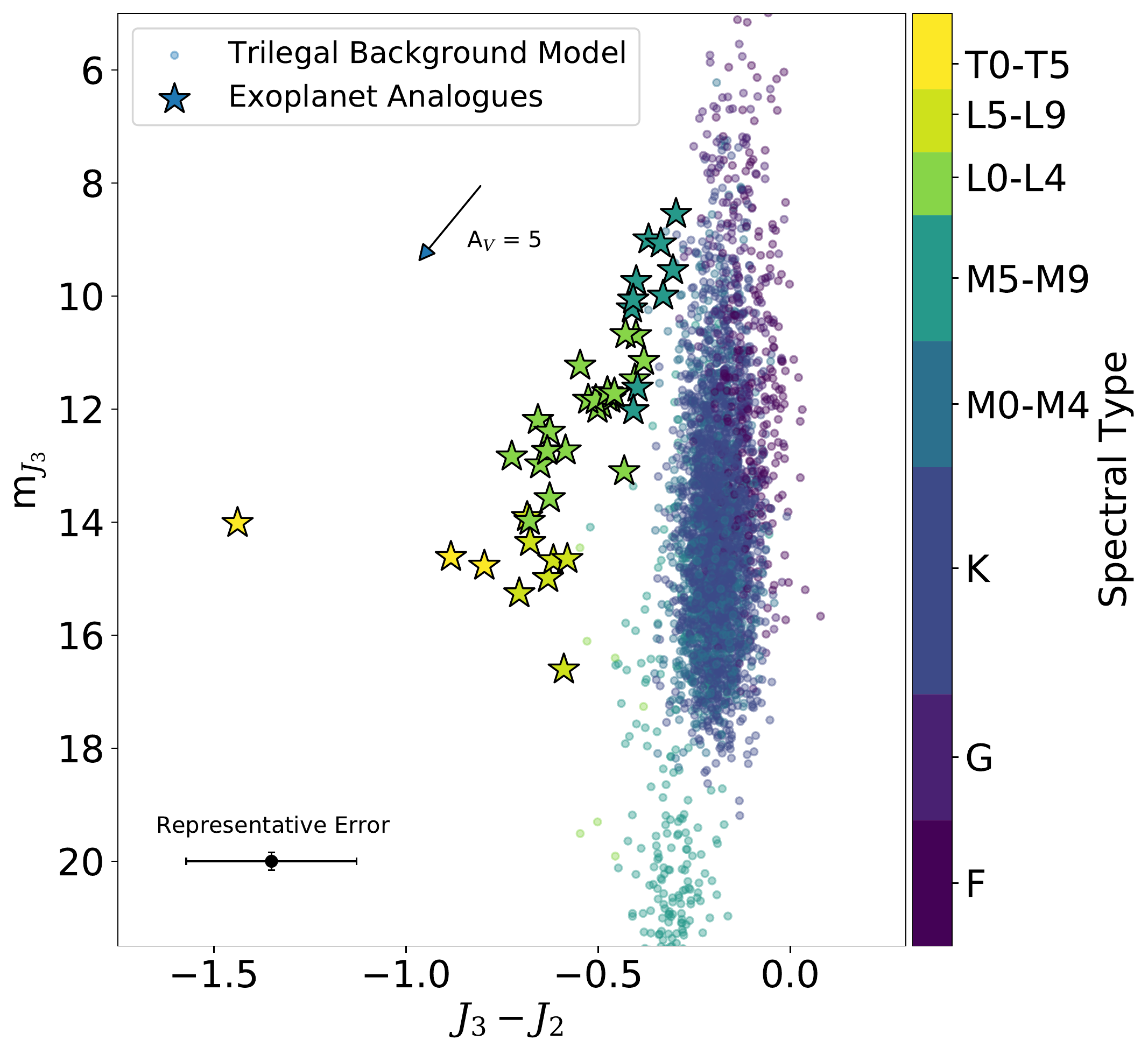}
    \caption{$J_{3}-J_{2}$ vs m$_{J_{3}}$ colour-magnitude diagram. Young, late-type exoplanet analogues (stars) are plotted, as well objects returned for Trilegal Galactic model along the $\beta$ Pictoris line-of-sight are plotted (circles), with the point colour indicating the spectral type of each object. Representative error bar from SHINE F150 shown in bottom left. The arrow plotted shows the $A_{\text{V}}$ = 5 extinction vector.}
    \label{fig:j3j2}
\end{figure}

Plotted in Figure \ref{fig:nix_colcol} are three datasets: first, in a gradient of orange squares, a sample of simulated stellar photometry with spectral types encompassing the spectral types expected in the field, from F0-M2. This population is reddened by $A_{\text{V}} = 0, 5, 10, 15$ and 20. We used spectral standards with these spectral types taken from the IRTF Spectral Library \citep{rayner09}, and performed synthetic photometry on each reddened spectra in the $K$-peak, IB2.42 and $H$2-cont filters. We fit a best fit line to this distribution, and show the 3$\sigma$ spread as dashed lines in Figure \ref{fig:nix_colcol}.

We then simulated a realistic background field population using the Trilegal Galactic models \citep{girardi12}. This simulation required a possible line-of-sight for the NIX survey. We chose $\beta$ Pictoris as the target line of sight, a stellar member of the $\beta$ Pictoris moving group \citep{zuckerman01,gagne18a}. We simulated a field of 1$\times$1 arcmin (the approximate field of view of NIX), centred on RA = 86.8$^{\circ}$, Dec = -51.1$^{\circ}$. We used the Kroupa initial mass function (IMF) \citep{kroupa01}, including binaries, which results in a population of $\sim$26,000 objects along the $\beta$ Pictoris line-of-sight. Each model object returned by the Trilegal model has an associated effective temperature. We converted these to spectral types using Mamajek's `Modern Mean Dwarf Stellar Color and Effective Temperature Sequence' \footnote{\hyperlink{}{https$://$www.pas.rochester.edu/$~$emamajek/EEM\_dwarf\_UBVIJHK\_colors\_Teff.txt}, accessed June 2021} \citep[described in part in ][]{pecaut13}. With a spectral type for each model object, we could then associate it to a spectral standard (using template with the same spectral types from the IRTF spectral library), and perform synthetic photometry in the three filters. The resulting population is plotted as the coloured circles in Figure \ref{fig:nix_colcol}, with the colour bar showing the spectral type of each object. As expected, objects with earlier spectral types occupy the same position as the simplified field model described above, but the Trilegal population also shows us the location of field mid-M and L objects on the colour-colour diagram, which are distinct from the earlier type objects.

The final population plotted on Figure \ref{fig:nix_colcol} is a selection of young exoplanet analogues. These range in spectral type from M5 - T5.5, and are all free floating objects that have been spectroscopically characterised. The sample of objects is described in detail in \citet{bonnefoy18}. Synthetic photometry was performed on the spectral data in the same way as above, and the colour bar in Figure \ref{fig:nix_colcol} again corresponds to the spectral type of these targets. \par

We can see a clear distinction between the reddened, background F0-M2 stars and the young exoplanet analogues on this colour-colour diagram. The optimal custom filter design motivated by this: the young exoplanet analogues that would be targets of a direct imaging survey lie in a (mostly) unique part of this parameter space, with a clear colour offset when compared to the F0-M2 background sequence. With this combination of filters, there is a still small amount of cross-over in the parameter space covered by field-age and young objects of the latest spectral types. Despite this, the $K$-peak filter will allow us to obtain a better understanding of the spectral type of many targets from the colours alone. Through simulated photometry, we calculated that a colour offset measured to 0.1 mag or better will be able to robustly distinguish young planetary-mass object candidates from background contaminants. This photometric precision should be very achievable with the NIX imager. As a result of this analysis, the final $K$-peak filter was manufactured with a central wavelength of 2.2$\mu$m and a width of 6\%.

\section{Possible Observing Strategies} \label{sec:surveys}

Our custom $K$-peak filter has the potential to be a powerful tool for identifying very low-mass candidate objects in imaging data. In this next section, we will demonstrate the capabilities of a NIX direct imaging survey. We consider multiple survey approaches, and discuss the potential performance of NIX in each case. We do not present fully-formed survey designs, as instrument commissioning is planned for 2022, and actual instrument performance will not be known until after commissioning

We consider two categories of observational strategies: `targeted' and `regional' surveys. The targeted approaches we consider below involve selecting specific host stars, based on prior knowledge of possible companions from other survey data. Conversely, a regional survey approach would target all host stars that meet certain selection criteria, and are members of a chosen region, such as a specific moving group. We will discuss the pros and cons of each observational strategy in Section \ref{sec:disc}.

\subsection{Targeted Surveys} \label{sec:surveys_targeted}

\subsubsection{Following up low-proper motion and crowded fields from the SHINE survey} \label{sec:surveys_crowded}

As discussed in Section \ref{sec:intro}, the first results from the ongoing SPHERE SHINE direct imaging survey have recently been published \citep{vigan21,langlois21,desidera21}. In \citet{langlois21}, candidate companions are identified using their proximity to their host star, and then plotted on colour-magnitude diagrams using various combinations of SPHERE narrowband filters. Where they lie in these colour magnitude diagrams can be used as an indicator of whether they are a low-mass companion to the imaged host star, or an interloping background contaminant. In Figure \ref{fig:j3j2}, we replicate one such colour-magnitude diagram using the $J$2, $J$3 dual filter. The two populations of objects shown are those described in Section \ref{sec:customfilter_design} and plotted on Figure \ref{fig:nix_colcol}: an average distribution of contaminant background objects from the Trilegal Galactic model of $\beta$ Pictoris, and the group of young exoplanet analogue objects described in \citet{bonnefoy18}, which have spectral types ranging from mid-M to mid-T. We can compare Figure \ref{fig:j3j2} directly to Figure \ref{fig:nix_colcol} to assess the diagnostic capabilities of the two filter sets, which both use the spectral shape of the targets to characterise them. 

In Figure \ref{fig:j3j2}, the background sequence lies on a well defined colour locus. The exoplanet analogue population begins to merge with this background sequence for the brightest objects, which includes some mid-late young Ms. In general the young L- and T- type objects and distinct from the background, with a few exceptions that might prove difficult to characterise if poor conditions lead to large photometric errors. A typical errorbar for the SHINE F150 sample is shown in the bottom right. This was calculated by obtaining the measured $J$2 and $J$3 errors in the published catalogue of SHINE F150 observations \citep{langlois21}, and taking the peak values of the full distributions of these errors. We then added these errors in quadrature to find the overall $J$2-$J$3 error. It should be noted that this error is likely an overestimation, as $J$2,$J$3 is a dual filter, with photometric observations obtained simultaneously. If we consider these representative errors for the young exoplanet analogues, we can see that it may not be possible to robustly distinguish M-L type objects from the background locus for some cases using only SPHERE photometry. Considering instead the colour-colour diagram shown in Figure \ref{fig:nix_colcol}, the distribution of exoplanet analogues is very spatially distinct when compared to the F-M background sequence, even if large photometric errors are considered (3$\sigma$ region indicated by dashed lines). As in the SPHERE colour-magnitude diagram, there are regions of overlap, but for populations of different ages rather than mixing of early and late type objects.

The NIX spectral shape technique could be used to critically confirm or refute candidate companions found previously in other imaging surveys, that are yet to be characterised. By considering the positions of targets in both the $J$2, $J$3 colour-magnitude diagram and the NIX colour-colour diagram, we will be able to estimate spectral types for previously observed objects with just further photometric observations, rejecting background K-M stars and dramatically reducing the intensiveness of follow-up time. 
Considering the simultaneous operation of both SPHERE and ERIS at the VLT, following-up SPHERE-observed objects with a different instrument may not often be a sensible approach. However, as demonstrated in Section \ref{sec:context_pm}, the spectral shape technique could prove incredibly useful for specific objects in fields with low proper motion. For targets with first-epoch SPHERE observations in such fields (for example, more distance star-forming regions such as Upper Scorpius that have been targeted by SHINE), second-epoch observations with NIX could add the additional photometric information required to loosely characterise prior to spectroscopic or astrometric follow-up.

\begin{figure*}
    \centering
    \includegraphics[width=\textwidth]{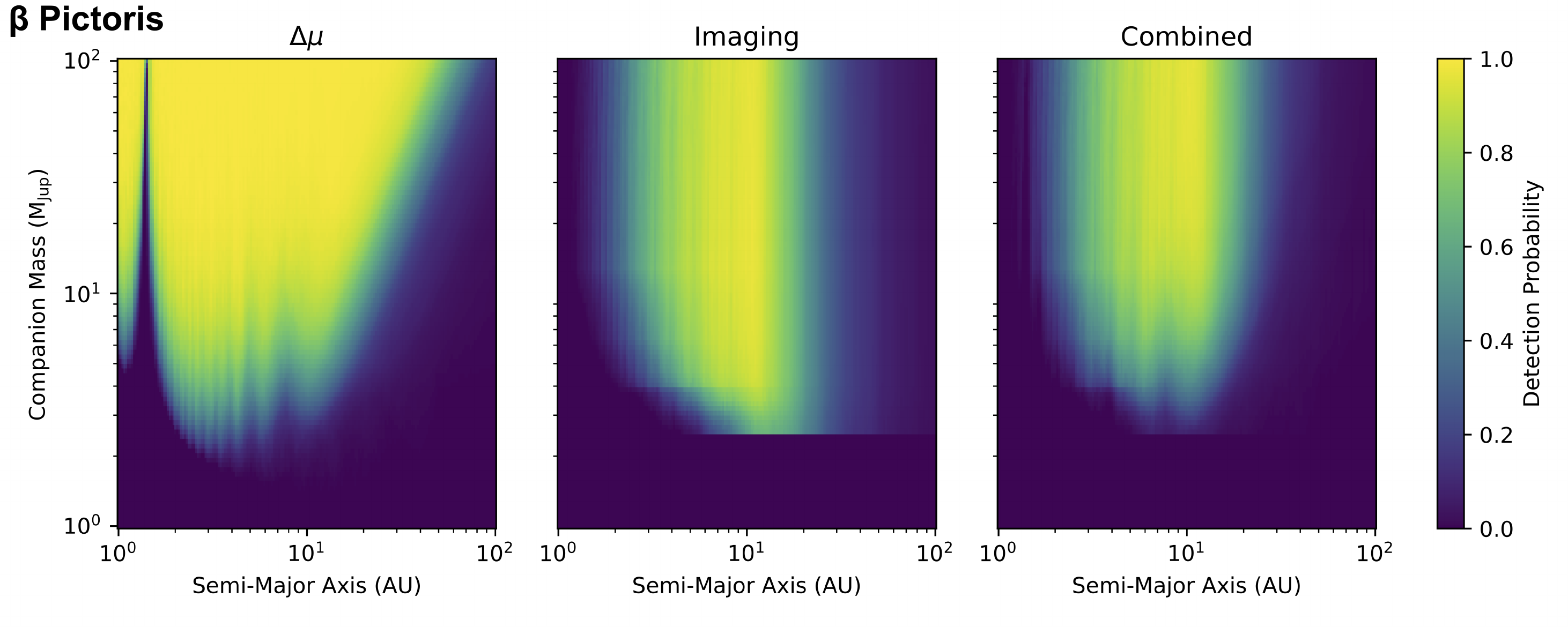}
    \includegraphics[width=\textwidth]{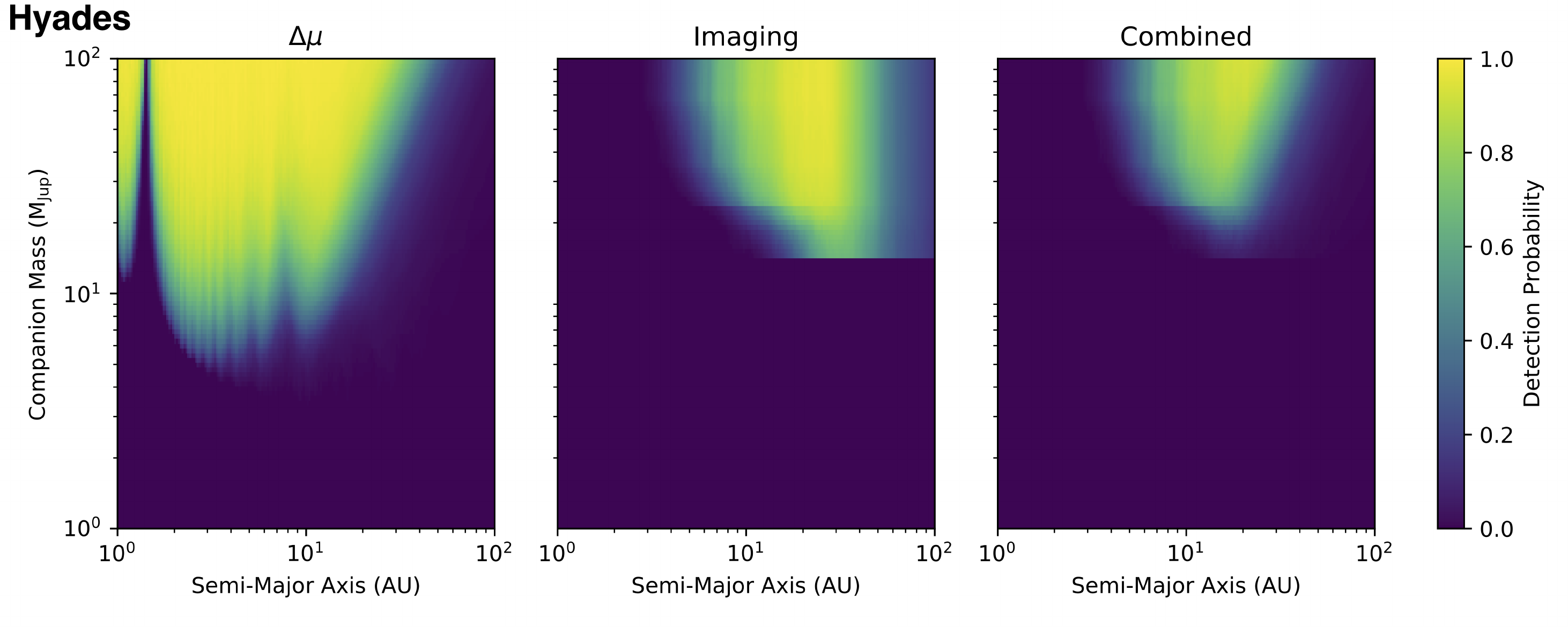}
    \caption{COPAINS simulation results for a 1 M$_\odot$ star in the $\beta$ Pictoris moving group (top) and Hyades cluster (bottom). Left: simulated systems that would be selected for observations by COPAINS, in planet mass--semi-major axis space. Colourbar shows fraction of systems at each grid point with $\ge$ 3$\sigma$-significance $\Delta\mu$ between TGAS and Gaia EDR3. The peak of 0\% astrometric detections at small separations corresponds to systems with orbital periods comparable to the timescale of Gaia EDR3, which thus have $\sim$null simulated $\Delta\mu$ values.
    Centre: Completeness in direct imaging observations of the same simulated systems with NIX. Colourbar shows the fraction companions in the drawn systems at each grid point that fall above the adopted sensitivity threshold for NIX for observations in 2022.
    Right: Combined completeness of the $\Delta\mu$ and imaging analyses from the left and middle panels. Colourbar shows the fraction of simulated companions that were successfully selected with COPAINS and detected in NIX observations.}
    \label{fig:COPAINS_results}
\end{figure*}

\subsubsection{Confirming $\Delta\mu$ selected candidates} \label{sec:surveys_deltamu}

Another possible observing strategy is to target host stars displaying anomalous proper motion trends. Expansive astrometric catalogues covering long time baselines allow us to examine the proper motions of millions of stars \citep[e.g][]{gaia21,vanleeuwen07}. Stars showing a discrepancy in proper motion between catalogues can be an indication of a hidden perturber affecting their motion, possibly a planetary companion. This technique facilitates a more informed choice of host stars to target: when they would usually be chosen based on mass, age, brightness etc, we can add the additional information of the possible presence of a companion, before any observations are undertaken. Multiple low mass companions have been detected using this method \citep[e.g][]{kervella19,currie21}. \par

COPAINS (Code for Orbital Parametrisation of Astrometrically Inferred New Systems; \citealp{fontanive2019}) is an innovative tool developed to use this technique, and identify previously undiscovered companions detectable via direct imaging, based on changes in stellar proper motions across multiple astrometric catalogues. We used COPAINS to estimate the range of systems that could be selected for their proper motion anomalies and be detectable with NIX ($\Delta\mu$ systems, where $\Delta\mu$ is the difference between the instantaneous velocity of a target and its true barycentric motion). We consider $\Delta\mu$ measurements between long-term proper motions from the Tycho-Gaia Astrometric Solution (TGAS; \citealp{michalik2015}) subset of the Gaia Data Release 1 catalogue \citep{gaia16b,gaia16a}, and short-term measurements from Gaia EDR3 \citep{gaia21}.

Figure \ref{fig:COPAINS_results} shows the results of $\Delta\mu$ analyses from COPAINS for typical targets of a direct imaging survey: a 1-M$_\odot$ star in the $\beta$ Pictoris moving group (top), with a parallax of 50~mas (distance of 20~pc) and proper motion of $\mu_\alpha = 5$~mas\,yr$^{-1}$ and $\mu_\delta = 80$~mas\,yr$^{-1}$, at an age of $24\pm3$~Myr \citep{bell2015}, and a 1-M$_\odot$ star in the Hyades stellar cluster (bottom), with a parallax of 21~mas (distance of 47~pc) and proper motion of $\mu_\alpha = 101$~mas\,yr$^{-1}$ and $\mu_\delta = -28$~mas\,yr$^{-1}$, at an age of 650~Myr \citep{lodieu19}. The left panels show the positions in the planet mass--semi-major axis space of companions that would show a significant change in proper motion between TGAS and Gaia EDR3, and would thus be selected with the COPAINS tool \citep{fontanive2019}. For each cell in the grid, $10^4$ random orbits were generated, adopting a uniform eccentricity distribution and drawing random inclinations and orbital phases, and the expected difference in proper motion between these two catalogues was calculated. The colourbar shows the fraction of these simulated systems in each point in the grid that have a $\Delta\mu$ significance of at least 3$\sigma$ between the TGAS and Gaia EDR3 catalogues, assuming combined uncertainties of 0.1~mas/yr in the proper motions (e.g., \citealp{brandt21}). 
The middle panels show the completeness in direct imaging observations with NIX, using a predicted contrast curve. As NIX is yet to be installed and commissioned on the VLT, there is no measured contrast curve from this specific instrument that can be used. Instead, we made use of the 3.9$\mu$m ($L'$-band) 5$\sigma$ contrast curve presented in \citet{otten17}. This contrast curve describes the on-sky performance of the vector apodising phase plate \citep[vAPP,][]{snik12,otten14} coronagraph installed on MagAO/Clio2 \citep{close10,close13,sivanandam06,morzinski14} at the {\it Magellan}/Clay telescope. Similar in design to the NIX grating vector apodising phase plate (gvAPP) coronagraph, it is the most up-to-date measured contrast curve suitable for synthetic NIX observations, before the installation of NIX on the VLT. As this contrast curve is calculated for a specific $L'$-band wavelength, it was scaled for use with the custom $K$-peak filter. The contrast actually achieved with NIX is likely to be comparable or better than the measured MagAO contrast curve. Most likely, scaling this contrast curve for use with the $K$-peak filter will be underestimating the contrast that will be measured with NIX, meaning the predictions reported in this paper are likely to be conservative. The predicted contrast curve was then converted into a mass limit using the AMES Cond evolutionary models \citep{allard2001} for the VLT/NaCo $Ks$ band, at the adopted distance and age for our typical 1~M$_\odot$ $\beta$ Pictoris moving group and Hyades stars. 

Using the same simulated systems as in the left panel of Figure \ref{fig:COPAINS_results}, we calculated the projected separation that would be observed at UT date 2023.0 for each companion, and checked whether it was detectable given the sensitivity limit. The colourbar again indicates the fraction of companions in each cell of the grid that falls above the estimated contrast curve. The right panels show the systems which would be selected with COPAINS and that would be detectable with ERIS, corresponding to the regions of high probabilities in both the left and middle panels. For a $\beta$ Pictoris star, this combined optimal range corresponds to separations of $\sim$3$-$15~AU with masses above $\sim$7$-$10~M$_\mathrm{Jup}$. For the Hyades, the slightly farther distance and older age of the cluster results in a poorer sensitivity to planetary-mass companions, especially on the imaging side, with a combined optimal range at $\sim$10$-$30~AU for substellar companions above $\sim$20$-$40~M$_\mathrm{Jup}$. However, as mentioned above, the predicted imaging contrast curve used here is likely to be rather pessimistic, and the overall range of companions amenable to a selection process like COPAINS is likely to extend to lower masses.
We compare these predicted performances to the other observing strategies in Section \ref{sec:disc}. We also note that given the strong dependence on the target's distance in the overlapping regions of sensitivity between astrometric analyses (affecting the limiting mass sensitivity) and imaging data (affecting the probed inner working angle), young associations like Upper Scorpius (145~pc; \citealp{preibisch08}) are not yet amenable to such informed selection methods to search for companions in the substellar regime.

\subsection{Regional surveys} \label{sec:surveys_blind}

\subsubsection{Young Star-forming regions} \label{sec:surveys_sfr}

\begin{figure*}
    \centering
    \includegraphics[width=\textwidth]{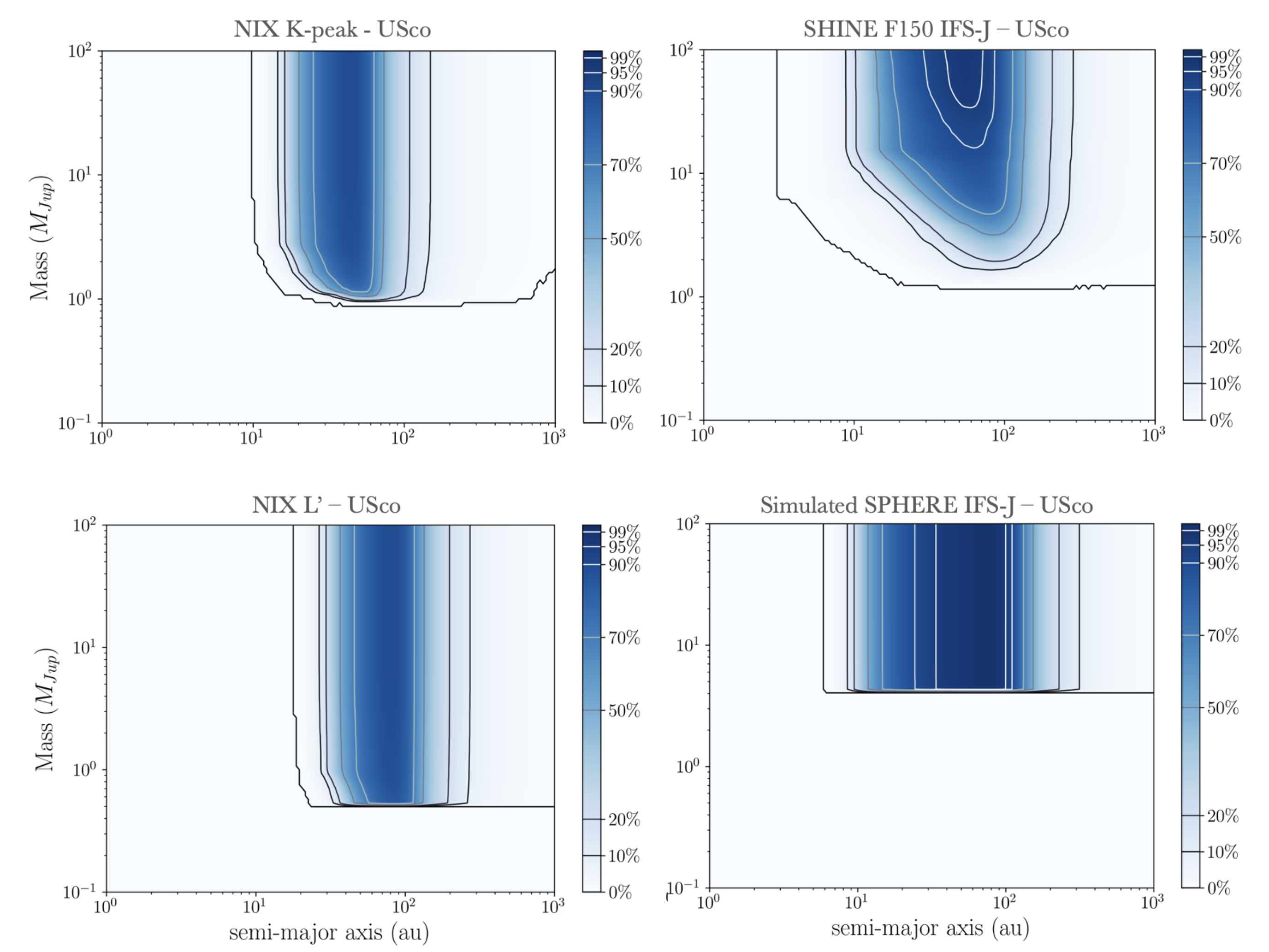}
    \caption{Median sensitivity maps for Upper Scorpius. Left: results from Exo-DMC simulations for the NIX $K$-peak (top) and $L'$ (bottom) filters. Right: comparable results for the SPHERE imager, using the same synthetic target list (bottom), and the real SHINE F150 Upper Scorpius targets (top). Colour indicates the percentage of companions detected with the corresponding mass and semi major axis. The contour levels are indicated in the colour-bar.}
    \label{fig:sensitivity_usco}
\end{figure*}

Next, we consider a survey approach that would focus on members of young star-forming regions. Such regions can contain very young stars, and the active star formation increases the possibility of observing planets and brown dwarfs either during or just after formation i.e at their brightest. As discussed in Section \ref{sec:context_archive}, star-forming regions have typically been less favourable targets in previous imaging surveys, primarily due to their distance and the resulting observational baseline required to confirm candidate companions via their proper motion. 

To analyse the suitability of stars within such a region for a NIX imaging survey, we can consider the sensitivity achievable for a specific survey. Exo-DMC \citep{bonavita20} is the latest (and first {\tt python}) rendition of MESS \citep[Multi-purpose Exoplanet Simulation System,][]{bonavita12}, a Monte Carlo tool for the statistical analysis of direct imaging survey results. 
In a similar fashion to its predecessors, the DMC combines information on the target stars with the instrument detection limits to estimate the probability of detection of a given synthetic planet population, ultimately generating detection probability maps. \par

For each star in a supplied sample, the DMC produces a grid of masses and physical separations of synthetic companions, then estimates the probability of detection given the provided detection limits. The default setup uses a flat distribution in log space for both the mass and semi-major axis but, similar to its predecessors, the DMC allows for a high level of flexibility in terms of possible assumptions on the synthetic planet population to be used for the determination of the detection probability.  
For each point in the mass/semi-major axis grid the DMC generates a fixed number of sets of orbital parameters. By default all the orbital parameters are uniformly distributed except for the eccentricity, which is generated using a Gaussian eccentricity distribution centred at $\mu =0$ with a width of $\sigma = 0.3$ (for positive values of eccentricity), following the approach by \cite{hogg10} \citep[see][for details]{bonavita13}. 
This allows for proper consideration of the effects of projection when estimating the detection probability using the contrast limits. The DMC in fact calculates the projected separations corresponding to each orbital set for all the values of the semi-major axis in the grid \citep[see][for a detailed description of the method used for the projection]{bonavita12}. This enables the estimation of the probability of each synthetic companion truly being in the instrument FoV and therefore being detected, given that the value of the mass is higher than the corresponding limiting mass.

Figure \ref{fig:sensitivity_usco} shows the results of the Exo-DMC simulations for the Upper Scorpius star forming region. Upper Scorpius is an ideal target for the NIX imager. In Section \ref{sec:context_archive}, we described an Upper Scorpius member list derived from the compilation presented in \citet{luhman18}. We use this for the remainder of the Upper Scorpius analysis, with the additional assumptions of a common age of 10 Myr \citep{pecaut16} and a common distance of 145 pc \citep{preibisch08} for all Upper Scorpius members in our final list (141 targets). 

Figure \ref{fig:sensitivity_usco} shows median-combined sensitivity plots for the $K$-peak filter (top-left), the standard $L'$-band (bottom-left) and, for comparison, SPHERE SHINE (right). Two results are shown for SPHERE SHINE: a synthetic sensitivity map (bottom-right), created using the target list above, and the SPHERE IFS $J$-band, and the sensitivity of real IFS $J$-band observations of Upper Scorpius in the SHINE F150 sample \citep[top-right,][]{vigan21}. Contour levels are indicated in the colourbar, with the minimum contour being 0\% in each case. In the following analyses, we will consider the best-case mass depths as dictated by the 10\% contour. 

The top- and bottom-left panels indicate that our NIX sensitivity is comparable in both filters for targets in Upper Scorpius, in both cases suggesting we will be most sensitive to planets from $\sim$ 30--100 AU, with the $K$-peak filter reaching masses of $\sim$1M$_\mathrm{Jup}$, and $L'$ likely able to detect less massive planets down to $\sim$0.5M$_\mathrm{Jup}$, but at typically wider separations. Comparing this to the sensitivity of the SPHERE/SHINE IFS $J$-band (upper-right), we can clearly see that the performance of NIX will be a improvement in a similar part of semi major axis/mass parameter space. The NIX filters offer a slightly narrower coverage in semi major axis, but far deeper coverage in mass: NIX observations of the same targets in Upper Scorpius would be sensitive down to $\approx$1M$_\mathrm{Jup}$, whereas SPHERE is generally most sensitive to planets $\textgreater$3M$_\mathrm{Jup}$. The SHINE F150 sensitivity map is, as expected, similar to the modelled map for the IFS $J$-band - but suggests a better mass sensitivity for some of the semi-major axis space covered. For the remaining regions we discuss in the work, we will only present the real SHINE F150 sensitivities. Despite these being derived from different target lists to the NIX results, they are a more meaningful comparison, as they demonstrate the true, achieved sensitivities of the SHINE survey in each region. 

\subsubsection{Young moving group members} \label{sec:surveys_gymg}

\begin{figure*}
    \centering
    \includegraphics[width=\textwidth]{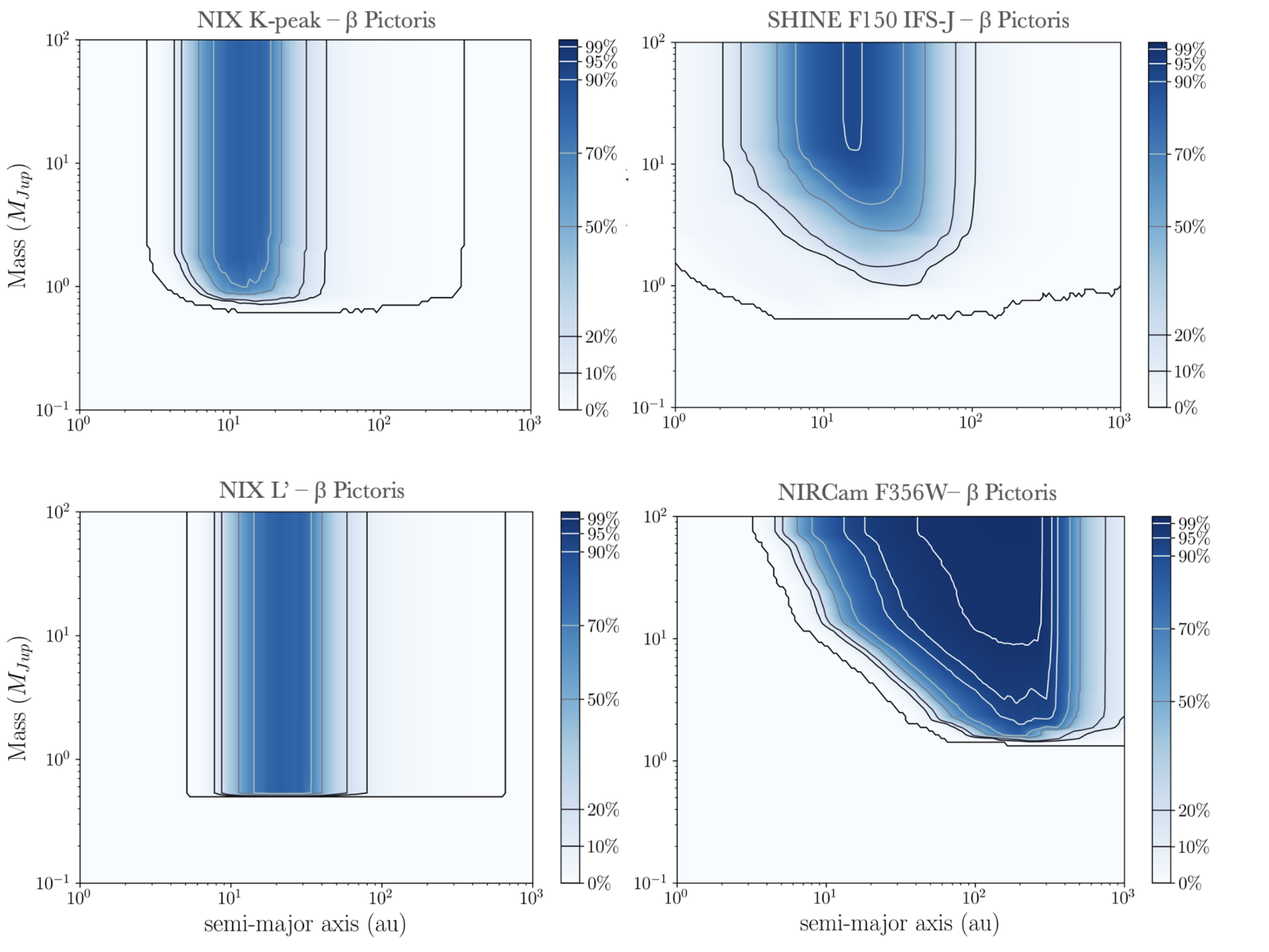}
    \caption{Median sensitivity maps for the $\beta$ Pictoris moving group. Left: results from Exo-DMC simulations for the NIX $K$-peak (top) and $L'$ (bottom) filters. Right: $\beta$ Pictoris sensitivity results using real SHINE F150 targets (top) and simulated JWST F356W results \citep[bottom][]{carter21}. Colour indicates the percentage of companions detected with the corresponding mass and semi major axis. The contour levels are indicated in the colour-bar.}
    \label{fig:sensitivity_bpic}
\end{figure*}

\begin{figure*}
    \centering
    \includegraphics[width=\textwidth]{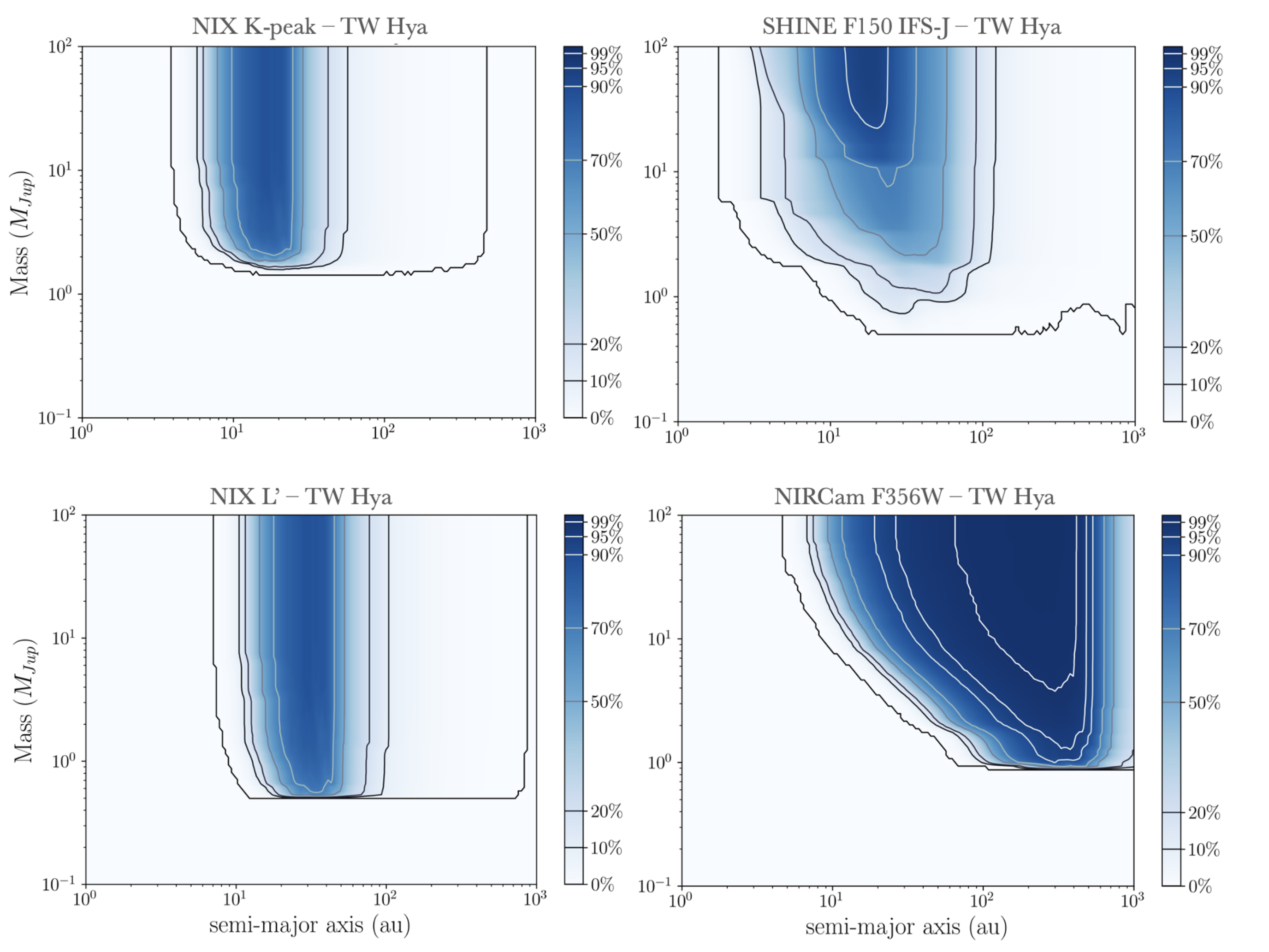}
    \caption{Median sensitivity maps for the TW Hya moving group. Left: results from Exo-DMC simulations for the NIX $K$-peak (top) and $L'$ (bottom) filters. Right: TW Hya sensitivity results using real SHINE F150 targets (top) and simulated JWST F356W results \citep[bottom][]{carter21}. Colour indicates the percentage of companions detected with the corresponding mass and semi major axis. The contour levels are indicated in the colour-bar.}
    \label{fig:sensitivity_twa}
\end{figure*}

As discussed in Section \ref{sec:context_archive}, large direct imaging surveys have often targeted samples of stars belonging to nearby young moving groups, due to both their proximity and youth. Observations of moving groups dominate the archival catalogue of imaging data when compared to the previously discussed star-forming regions for this reason. In Section \ref{sec:context_archive}, we presented two lists of moving group members, for the $\beta$ Pictoris and TW Hya moving groups \citep{gagne18a,carter21}. We will use these member lists for the remainder of this analysis.

Exo-DMC simulations (as described above) were performed for the compilation of $\beta$ Pictoris members taken from \citet{carter21}, with results shown in Figure \ref{fig:sensitivity_bpic}. The sensitivities for the individual stars were again median combined. The SPHERE sensitivities shown here are again the real results for $\beta$ Pictoris members observed in the SHINE F150 sample. For $\beta$ Pictoris moving group stars, the sensitivity in semi-major axis across the two NIX filters is $\sim$5-50 AU, reaching closer into the target stars than observations of Upper Scorpius (Figure \ref{fig:sensitivity_usco}). In terms of potential mass depth that could be achieved by targeting stars in $\beta$ Pictoris with NIX, planetary companions with masses $\sim$1M$_\mathrm{Jup}$ could be detected in both the $K$-peak and $L'$ filters. Comparing this predicted performance to the actual performance of SHINE F150 IFS $J$-band in $\beta$ Pictoris, we can see that NIX will be sensitive to slightly less massive planets in both filters, but at similar separations to those probed by SPHERE.

Figure \ref{fig:sensitivity_twa} also shows the results from Exo-DMC simulations for the TW Hya members taken from \citet{carter21}. Comparing these sensitivity maps to Figure \ref{fig:sensitivity_bpic}, the performance of NIX and SPHERE for the two moving groups is clearly very similar, as expected due to their comparable ages and distances. There are some notable distinctions: the most sensitive area of parameter space is shifted to slightly wider separations for TW Hya members when compared to $\beta$ Pictoris, because of its slightly increased distance. Additionally, the minimum detectable mass in $K$-peak is higher in TW Hya, reaching a depth of $\sim$2M$_\mathrm{Jup}$, while the potential mass sensitivity in $L'$ is essentially the same. Comparing the potential performance of NIX to SPHERE for TW Hya, we again see a similar result in mass sensitivity in the $K$-peak filter, and a considerable improvement for observations using the longer wavelength $L'$ band.

Having extensively compared the predicted performance of NIX to SPHERE, we are also interested in its potential when compared to another future tool that will likely prove very successful in the field of exoplanet science. In both Figures \ref{fig:sensitivity_bpic} and \ref{fig:sensitivity_twa}, we present a fourth panel, which compares the sensitivity performances of NIX and SPHERE for each moving group to the upcoming {\it James Webb Space Telescope} \citep[JWST;][]{gardner06}. The instrument chosen for comparison is the Near-InfraRed Camera \citep[NIRCam;][]{rieke05}, which has a similar long-wavelength capability (0.6--5$\mu$m) and angular resolution as NIX. Although not primarily designed as a coronagraphic imaging instrument, NIRCam will regardless be used for giant planet coronagraphic imaging, and the F356W filter is directly comparable to the NIX $L'$-band. 

In the bottom-right panels of Figures \ref{fig:sensitivity_bpic} and \ref{fig:sensitivity_twa}, we present median sensitivity maps for $\beta$ Pictoris and TW Hya in the NIRCam F356W filter. These use the results of \citet{carter21}, who simulate the mass sensitivity limits of JWST coronagraphy for four NIRCam and MIRI filters. \citet{carter21} use the MASK335R round coronagraphic mask for all of their simulations. For both moving groups, NIRCam F356W is most sensitive to planets at wider separations that the NIX or SPHERE results, reaching the lowest depth in mass (approximately 1M$_\mathrm{Jup}$) from $\sim$100 AU onwards.

\section{Discussion} \label{sec:disc}

\subsection{Merits of each survey approach} \label{sec:disc_merits}

In this work, we have presented four possible approaches for a future large-scale imaging survey with ERIS NIX. The focus of the design of these options was to maximise the unique capabilities of the custom $K$-peak filter and the long wavelength options of NIX. In summary, these survey approaches are: 
\begin{enumerate}
    \item  {Obtaining second epoch observations of stars with unconfirmed candidate companions in low proper motion fields that were previously observed in the SPHERE SHINE campaign.} Combining NIX data with existing SPHERE photometry could provide the additional information necessary to characterise and confirm (or refute) previously identified candidates.
    \item Using the COPAINS tool to choose imaging targets based on $\Delta\mu$ trends between different astrometric surveys. The spectral shape technique would be useful for characterisation, and the target list would be informed by COPAINS results.
    \item A survey of nearby young star forming regions. This approach is similar to what has been done by previous collaborations - using a target list informed by likely membership of a region, which itself is chosen using distance and age. The unique aspect of such a survey with NIX is that the custom $K$-peak filter can rule out significant numbers of background interlopers without the necessity of time-consuming proper-motion followup observations, opening up the possibility of surveying more distant star forming regions that have been relatively neglected. Additionally, the deep sensitivity of $L'$ will likely prove very useful in extending our sensitivity to considerably lower planet masses in such regions, and allow for studies of protoplanets in circumstellar disks.
    \item A survey of young moving group stars. As with the above approach, this option is similar to past surveys, as young moving groups have been extensively targeted by direct imaging . Here, the custom $K$-peak filter is less uniquely useful, since proper motion follow up is not as time intensive. Instead, the long wavelength capabilities of NIX could lead to new detections in unexplored parts of parameter space.
\end{enumerate}

\par

\begin{figure*}
    \centering
    \includegraphics[width=\textwidth]{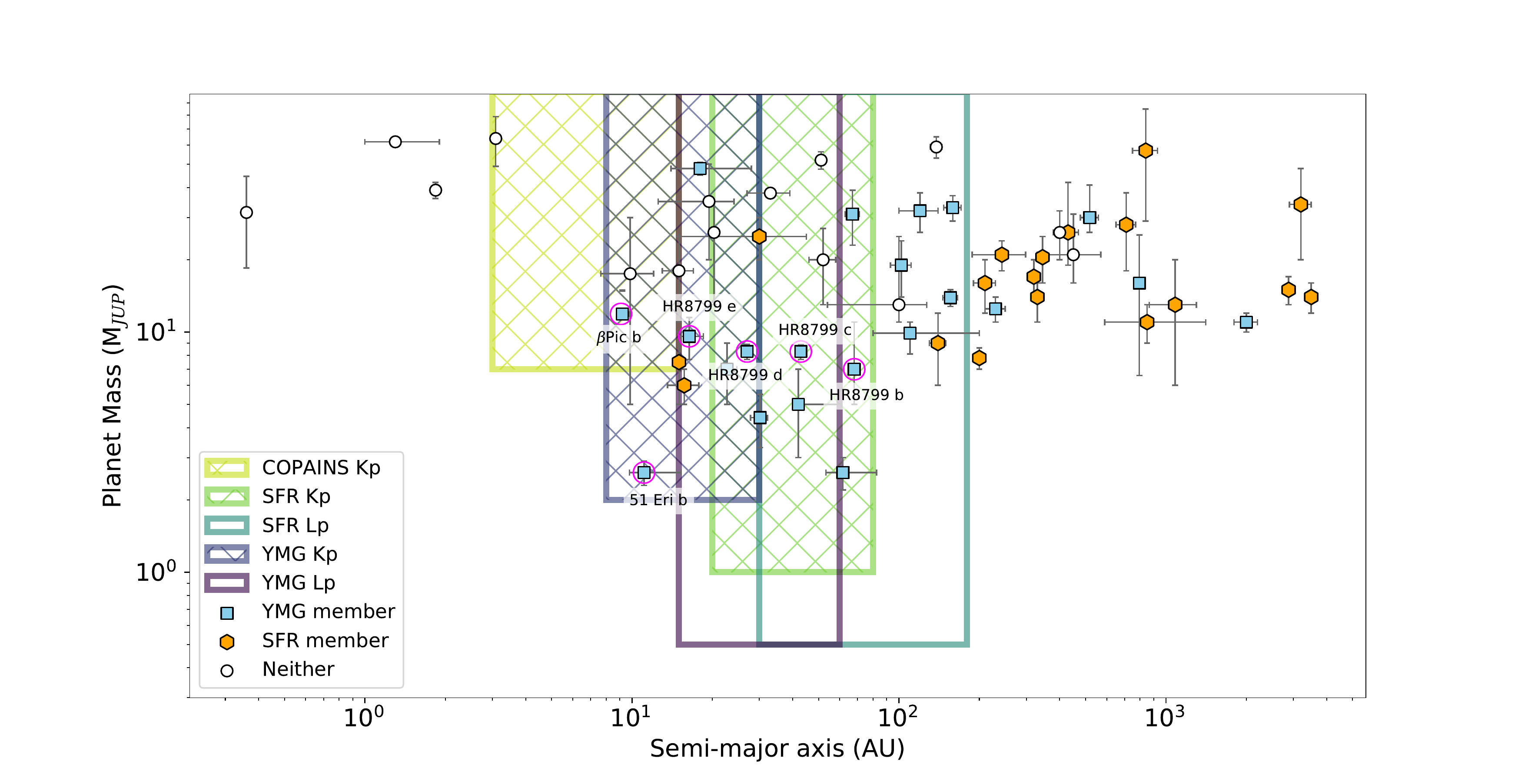}
    \caption{Planet mass vs semi-major axis. Plotted are planets and brown dwarfs detected using direct imaging (exoplanet.eu). Colour-coding of points indicates members of young moving groups (blue) or star forming regions (orange). Rectangular regions show the possible sensitivities for each survey approach. Cross-hatched areas indicate NIX simulations using the $K$-peak filter, areas with no fill are NIX simulations using the $L'$ filter.}
    \label{fig:di_space}
\end{figure*}

It is clear that the $K$-peak filter is specifically useful for most of the survey designs that we have explored in this paper, and also that the $L'$ capability of NIX will be extremely valuable. Figure \ref{fig:di_space} compares the regions of semi-major axis/ planet-mass space probed by three  of the approaches discussed above. In this Figure we show the sensitivities of each approach in either $K$-peak only, or $K$-peak and $L'$. We also plot a catalogue of known companions, including any planets and brown dwarfs with constrained planet masses and semi-major axes that are labelled as direct imaging detections\footnote{Taken from \hyperlink{}{exoplanet.eu}}.

First, we consider the planet mass--semi-major axis sensitivity predicted by the COPAINS simulations, for a typical $\beta$ Pictoris moving group star. In Section \ref{sec:surveys_deltamu}, we presented a sensitivity map generated using an approximate contrast curve for $K$-peak ($K$-peak results in Figure \ref{fig:di_space} are indicated by a cross-hatched shading). This showed that the highest probability region for selection by COPAINS and detection by NIX is 3--15 AU and 7--100 M$_\mathrm{Jup}$ (the highest mass considered), highlighted in lime green in Figure \ref{fig:di_space}. The planets we would be sensitive to using target selection informed by COPAINS for the nearby $\beta$ Pictoris moving group populate the closest semi-major axes of any of the survey approaches considered here, and reach into an area of parameter space where very few planets have thus-far been discovered via direct imaging. Furthermore, with all subsequent GAIA data releases, the sensitivity of this technique will continue to improve. Comparing the mass sensitivity predicted by COPAINS derived using Gaia DR2 vs eDR3, we already see an improvement of a few Jupiter masses when using the latest data release. However, it should be noted that the remaining yield from targeting 1-M$_{\odot}$ stars at $\sim$20 pc in $\beta$ Pictoris is likely low, due to the small number of these stars in this moving group, and the frequency with which they have already been observed. A region with reasonably similar properties that has been less favoured historically would be a good alternative target.

Next, we consider targeting young, nearby star forming regions, or young moving group stars. We explored Upper Scorpius as a possible star-forming region to target with a direct imaging survey. Exo-DMC simulation results for Upper Scorpius indicate comparable sensitivity in $K$-peak and $L'$ - with the most sensitive region being for planets between 20--80 AU for $K$-peak and 30--180 AU in $L'$. The $K$-peak and $L'$ sensitivities are highlighted in Figure \ref{fig:di_space}. In $K$-peak we reach planet masses of $\gtrsim$1M$_\mathrm{Jup}$, but the predicted performance in $L'$ is extremely encouraging, suggesting that masses down to $\sim$0.5M$_\mathrm{Jup}$ could be reached - the deepest mass limit of any of the survey approaches considered in this work. It should be noted that $\sim$0.5M$_\mathrm{Jup}$ is the lowest mass covered by the evolutionary models used in Exo-DMC, and as a result impose an artificial mass cut-off.

We also ran simulations of observations for two possible moving group targets: $\beta$ Pictoris (multiple lines of sight, compared to a singular target for the $\Delta\mu$ analysis in Section \ref{sec:surveys_deltamu}) and TW Hya, as discussed in Section \ref{sec:surveys_gymg}. We present the regions of high detection probability for the TW Hya moving group in Figure \ref{fig:di_space}, to allow comparison between a moving group target and a young star forming region. Here we can see that the $K$-peak sensitivity (again indicated by the cross-hatched area), probes closer semi-major axes than the results for Upper Scorpius (8--30 AU), but does not reach as deep in mass, with the minimum likely detectable mass $\approx$2 M$_\mathrm{Jup}$. In contrast, the result for $L'$ sensitivity for TW Hya moving group stars are very similar to the performance predicted for Upper Scorpius targets, again reaching a minimum detectable mass of $\sim$0.5M$_\mathrm{Jup}$. A survey approach using the $L'$ filter to survey TW Hya, or both $K$-peak and $L'$ to explore Upper Scorpius, would likely be sensitive to companions with masses \textless 2 M$_\mathrm{Jup}$, where very few directly-imaged planets have been discovered to date.

The conclusions that we can draw from Figure \ref{fig:di_space} are three-fold: first, independent of the survey approach chosen, NIX will be the most sensitive to both areas of parameter space where many planets have been directly imaged previously, and areas where little is known about the true direct imaging planet population. Secondly, using the COPAINS tool to inform a target list could allow us to probe a sparsely populated region of parameter space, with planetary companions close to their host stars. Thirdly, targeting more distant star-forming regions in general will likely lead to better depth in planet-mass than observing nearby young moving group stars, with the notable exception of an $L'$-specific survey of a nearby young moving group. 

\subsection{Planet Populations} \label{sec:disc_jwst}

In the previous section, we investigated the potential performance of a NIX survey, using the current population of direct-imaging discovered planets as a guide for what we might find. This approach does not give us the full picture: as we show in Figure \ref{fig:di_space}, survey detection space is limited by instrument design and performance. Consequently, the planets that have been discovered to date do not represent the distribution of planets as a whole. To understand the full distribution of planets in the semi-major axis-planet mass parameter space, we must use planet population models.

As discussed in Section \ref{sec:intro}, our understanding of the underlying planet population has evolved dramatically in recent years. Prior to the publication of recent results from the latest generation of surveys \citep{nielsen19,vigan21,desidera21,langlois21}, direct imaging searches were informed by predicted detection yields derived from the distribution of radial velocity planets. Studies such as \citet{cumming08} found that the detected radial velocity planets (typically at $\lesssim$3 AU) followed a rising power law in mass and orbital period, thus predicting many giant planets at wide separations from their host stars. Many direct imaging surveys \citep[including][]{lafreniere08,heinze10,macintosh14} were planned and interpreted using this assumption, as it was the best available at the time. Lower than expected yields from many such surveys confirmed suspicions that simply extending the RV power law did not accurately describe the giant planet population at wide separations.

We now have a better understanding of where we expect to find planets in a direct imaging survey. \citet{fernandes19} published a ground breaking study, using planets discovered by the {\it Kepler} telescope and radial velocity instruments \citep{mayor11} to derive a turn-over in planet occurrence rates at 2--3 AU. This model agreed with data from previous surveys \citep[e.g][]{bowler16,galicher16}, suggesting that giant planets on wide orbits were rarer than previously thought. However, the occurrence of planets beyond 3 AU was not well constrained by this study, as RV and transit data doesn't typically extend to long enough orbital periods to cover these separations. \citet{meyer18} considered the planets hosted by M-dwarf stars, and generated a fit to the population between 0.07 and 400 AU. They did this by using different datasets for different separations, including microlensing detections for 0.5-10 AU \citep{cassan12}, and direct imaging detections for separations $>10$ AU. Using a variety of datasets from different detection methods enabled them to optimise a log-normal function which best describes the data, and find a peak in planet frequency around M-dwarfs at 2.8 AU. However, the limitations of each detection technique must be considered: e.g. as microlensing targets primarily low-mass host stars, this technique could not easily be expanded beyond only M-dwarf stars.

\citet{fulton21} recently published a study of the planet population using data from the California Legacy Survey \citep{rosenthal21}, which has been observing the same sample of 719 nearby stars stars for $\sim$30 years. These stars range in spectral type from F-M stars, a much broader scope than those considered by \citet{meyer18}. The long observational baseline provided by this set of observations allowed them to place tight constraints on the giant planet occurrence rate, finding a turnover at $\sim$3.6 AU, and a well constrained downward slope out to $\sim$10 AU.

\begin{figure*}
    \centering
    \includegraphics[width=\textwidth]{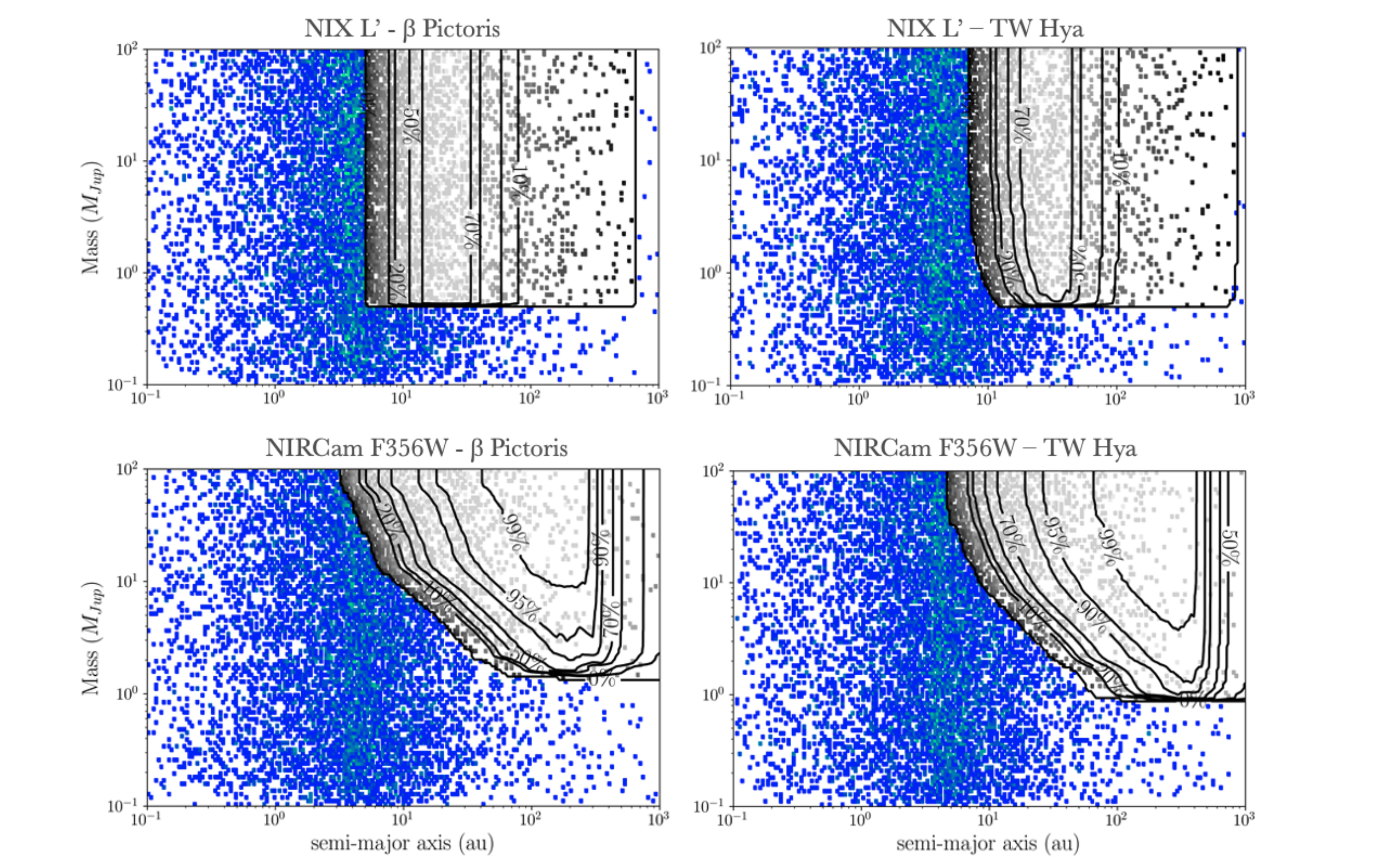}
    \caption{Instrument sensitivity comparison with a modelled planet population. Blue squares show the distribution of planets returned by the model described in \citet{fulton21}. In each panel, yellow/green overlay shows the sensitive area of parameter space for each instrument and region. Left: results for a survey $\beta$ Pictoris. Right: results for a survey of TW Hya. In both cases, using: NIX $L'$ (top) and NIRCam F356W \citep[bottom, limits from][]{carter21}}
    \label{fig:pops}
\end{figure*}

Figure \ref{fig:pops} uses the planet populations predicted by \citet{fulton21} to assess the sensitivity of a NIX survey targeting the two young moving groups discussed here ($\beta$ Pictoris and TW Hya, left and right panels, respectively) in the context of the predicted distribution of planets across the full parameter space. The distribution plotted as blue squares are the planets predicted by the \citet{fulton21} planet population between $0.1$M$_\mathrm{Jup} \leq M_{\text{P}} \leq 100$M$_\mathrm{Jup}$ and $0.1$AU$ \leq a \leq 1000$AU. Overlayed are the detection maps for $\beta$ Pictoris (left) and TW Hya (right) in the NIX $L'$-band and JWST F356W filter, taken from the plots shown in Figures \ref{fig:sensitivity_bpic} and \ref{fig:sensitivity_twa}. The labelled contours indicate varying levels of sensitivities.

As we discussed in Section \ref{sec:surveys}, Figure \ref{fig:pops} demonstrates that in the $L$-band wavelength range, NIRCam and NIX will be sensitive to essentially the same planet star separations, but NIX will be able to go deeper in mass at closer planet-star separations for both $\beta$ Pictoris and TW Hya. Based on the results of \citet{fulton21}, neither instrument is likely to be sampling the peak of the planet populations (at around 4AU) in $L$-band, but NIX will push closer to the peak for all planet masses considered here around stars in $\beta$ Pictoris. Regardless of the exact location of the peak of the distribution, there are many objects predicted in the tail of wider separation objects that both NIX and NIRCam will be sensitive to in $L$-band.

\section{Conclusions} \label{sec:concl}

We have presented the first details of a custom 2.2µm filter, the $K-$peak filter, that will be available on the NIX imager on the VLT. We aim to use this custom filter, and the spectral shape technique, to optimise a direct imaging survey for giant planets and brown dwarfs. We have also described four possible survey approaches, that aim to optimise survey yield and follow-up time requirements. The main takeaways from this analysis of observing strategies are:

\begin{itemize}
    \item NIX could be used to provide second epoch observations for some of the remaining candidates left at the end of the SPHERE SHINE survey.
    \item A survey informed by anomalous proper motion trends using the $K$-peak filter could prove fruitful in exploring the small projected separation part of parameter space.
    \item A survey targeting a young star forming region, such as Upper Scorpius, will likely result in a higher yield than a survey focusing on young moving groups. Firstly, moving groups have already been extensively surveyed, and for the two examples considered in this work the likely NIX contrasts are not better than those achieved by SPHERE. Secondly, we reach better mass sensitivity for star forming regions with NIX than SPHERE, and with the K-peak filter, we are less dependent on common proper motion confirmation follow-up to confirm (or reject) candidate companions.
\end{itemize}

Once on-sky, NIX will be a competitive instrument for direct imaging surveys. By considering possible observing strategies in advance, we aim to identify the optimal approach to maximise observing time efficiency and survey yield. The numerous large scale imaging surveys that have been completed to date or are currently underway are invaluable when planning future observations, as are analyses of planet populations. We aim to use the wealth of knowledge from recent years of direct imaging research to design a successful survey that will optimise both the detection and confirmation of candidate exoplanet companions.

\section*{Acknowledgements}

CF acknowledges support from the Center for Space and Habitability (CSH). This work has been carried out within the framework of the NCCR PlanetS supported by the Swiss National Science Foundation.
This work benefited from the support of the project FRAME ANR-20-CE31-0012 of the French National Research Agency (ANR).
This research has benefited from the SpeX Prism Spectral Libraries, maintained by Adam Burgasser at \url{http://pono.ucsd.edu/~adam/browndwarfs/spexprism/}.\\
This research has made use of the SIMBAD database, operated at CDS, Strasbourg, France, and NASA's Astrophysics Data System.\\

\section*{Data Availability}

The data underlying this article will be shared on reasonable request to the corresponding author.



\bibliographystyle{mnras}
\bibliography{paper.bib} 





\bsp	
\label{lastpage}
\end{document}